\documentstyle[11pt,psfig,aaspp4]{article} 
\def\etal{{\it et al. }}
\def\kms{km~s$^{-1}$~}
\def\nrin{374~}
\def\nin+{584~}

\def\ntot{782~}
\def\slop{-7.68~}

\def\W50{W$_{50}$~}
\received{Someday 1999}
\accepted{Someday 1999}
\journalid{337}{Someday 1999}
\articleid{11}{14}

\slugcomment{To appear in {\it Astronomical Journal}}

\begin{document}
\hskip 3.5in{Version 1.6 \hskip 10pt 14 Oct 1996}
\title{The I--Band Tully--Fisher Relation for Cluster Galaxies: 
A Template Relation, its Scatter and Bias Corrections}

\author {Riccardo Giovanelli, Martha P. Haynes, Terry Herter and Nicole P. Vogt}
\affil{Center for Radiophysics and Space Research
and National Astronomy and Ionosphere Center\altaffilmark{1},
Cornell University, Ithaca, NY 14853}

\author {Luiz N. da Costa}
\affil{European Southern Observatory, Karl--Schwarzschild--Str. 2, D--85748
Garching b. M\"unchen, Germany and Observatorio Nacional, Rio de Janeiro,
Brazil}

\author {Wolfram Freudling}
\affil{Space Telescope--European Coordinating Facility and European Southern 
Observatory, Karl--Schwarzschild--Str. 2, D--85748 Garching b. M\"unchen, 
Germany}

\author {John J. Salzer}
\affil{Astronomy Dept., Wesleyan University, Middletown, CT 06459}

\author {Gary Wegner}
\affil{Dept. of Physics and Astronomy, Dartmouth College, Hanover, 
NH 03755}

\altaffiltext{1}{The National Astronomy and Ionosphere Center is
operated by Cornell University under a cooperative agreement with the
National Science Foundation.}

\hsize 6.5 truein
\begin{abstract}

Infrared I band photometry and velocity widths for galaxies in 24 clusters,
with radial velocities between 1,000 and 10,000 \kms,
are used to construct a template Tully--Fisher (TF) relation. The sources 
of scatter in the TF diagram are analyzed in detail; it is shown that the 
common practice of referring to a single figure of TF scatter is incorrect
and can lead to erroneous bias corrections. Biases resulting from sample
incompleteness, catalog inaccuracies, cluster size and other sources, as 
well as dependences of TF parameters on morphological type and local 
environment, are discussed and appropriate corrections are obtained. 
A template TF relation is constructed by combining the data from the 24 
clusters, and kinematic cluster offsets from a putative reference 
frame which well approximates null velocity with respect to the cosmic 
microwave background, are obtained.

\end{abstract}


\section{Introduction}

This is the seventh paper of a series based on spectroscopy and I band 
photometry of spiral galaxies, obtained with the purpose of improving 
the calibration of the Tully--Fisher (TF) relation (Tully and Fisher 
1977) and our understanding of the peculiar velocity field in the
local universe. Previous papers in the series investigated the 
adequacy of internal extinction corrections for spiral disks (Giovanelli 
\etal 1994 and Giovanelli \etal 1995a, hereafter respectively
referred to as Paper I and Paper II); developed techniques for the
correction of selection biases in the determination of peculiar
velocities of field galaxies (Freudling \etal 1995, hereafter
Paper III); tested the evidence for large scale bulk flows (Giovanelli 
\etal 1996a; hereafter paper IV); presented a reconstruction of the
density field (da Costa \etal 1996; hereafter paper V); and provided 
a data set of galaxy parameters in cluster 
fields (Giovanelli \etal 1996b, hereafter Paper VI). The analysis
described here is based on the data presented in Paper VI,
which appears as a companion paper in this issue. Preliminary
reports of this work have appeared in Giovanelli 
\etal (1995b), which are superseded by this paper. The motivation
for this study is given in the introduction of Paper VI.

In Paper VI we present data for \ntot spiral galaxies in the fields
of 24 clusters in both hemispheres, with systemic velocities spread
between $cz \sim 10^3$ \kms and $cz \sim 10^4$ \kms, previously and 
throughout this paper referred to as the SCI sample . The tabulated
parameters were derived from our own observations or were collated from 
other sources in the public domain. 
The treatment of the data was carried out in such a way as to produce
a homogeneous set of TF parameters, which appear in Table 2 of Paper VI.

Cluster membership was assigned to each galaxy in the sample on the
basis of criteria somewhat more stringent than generally adopted in
previous TF work. Galaxies associated with each cluster were separated
into two classes: (i) an {\bf in} sample includes galaxies that are very
likely to be cluster members, on the basis of their sky and redshift
coordinates; (ii) a class of ``peripheral'' objects 
includes objects having a redshift quite close to the systemic one
of the cluster, but sufficiently removed spatially from the cluster center
so that a membership assignment cannot be reliably made. The combination
of {\bf in} objects and peripheral objects is referred to as the 
{\bf in+} sample. The global {\bf in} sample of the 24 clusters
includes \nrin galaxies, of which 360 are deemed good candidates
for TF work, while the {\bf in+} sample includes \nin+
objects, of which 555 are used for TF work. The remaining 198 galaxies 
in paper VI are either foreground 
or background objects, or members of groups/clusters inadequately
sampled for the purposes of this study. Sample assignments and 
justifications for exclusion are given in Table 2 of Paper VI.

Here we obtain TF fits to each of the 24 clusters and combine the data to
produce a template TF relation. Such a  template relation will be adopted 
in the analysis of the peculiar velocity field of clusters, as well as 
that of an all sky sample of field spiral
galaxies (the SFI sample), which will be presented elsewhere.

In section 2 we analyze in detail the characteristics of the TF error 
budget, while in section 3 we discuss morphological type differences
and the important issue of bias. In section 4 we describe the procedures
adopted in obtaining TF fits to the data, while in section 5 
we present the TF relations fit separately to each of the 24 clusters.
In section 6 we combine the cluster samples and obtain a global template.
In section 7 we investigate the residuals with respect to that template. 
The results are summarized in section 8. In the next paper of this series
(Giovanelli et al 1996c: paper VIII) we analyze the cluster motions with 
respect to the reference frame defined by the cluster set.

Throughout this paper, we will often refer to the absolute magnitude 
variable as

$$y = M + 5\log h \eqno(1)$$

\noindent
where $h=H_\circ/(100$ \kms Mpc$^{-1}$) and $M$ is the I band
absolute magnitude of a galaxy. Absolute magnitudes are 
tabulated in Paper VI on the assumptions that: (a) galaxies in
the {\bf in} samples are at the distance indicated by the systemic
velocity of the cluster measured in the Cosmic Microwave Background
(CMB) reference frame; this assumption implies that the clusters are 
at rest with respect to the CMB; (b) peripheral galaxies that complete
the {\bf in+} samples are at the distance indicated by each galaxy's
radial velocity. The assumption that individual clusters are at rest 
will in general be incorrect; magnitude offsets that take into account
cluster motions need to be applied in order to obtain correct values
of $y$. We give an estimate of such offsets in section 6.

We shall also define a velocity width variable

$$x = \log W - 2.5 \eqno(2)$$

\noindent
which will equal zero near $W = 300$ \kms, where $W$ is an estimate
of twice the maximum rotational velocity of a galaxy. This convention 
is dictated by the practice of deriving $W$ from Doppler--broadened
single--dish 21 cm line global profiles of galaxies. The relationship
between the measured linewidth and the rotational speed of the galaxy
is discussed in paper VI and, in further detail, by Haynes et al. (1997).

\section{Error Budget of the TF Relation}

The process of obtaining a template TF relation by combining several 
cluster TF diagrams requires an understanding of the characteristics of 
the error budget. This is necessary in the estimate of the amplitude of
the corrections which take into account the effects of known biases 
affecting the process. We thus review those characteristics before the 
procedures and corrections applied to obtain a template relation are 
discussed.

In this section, we shall then run ahead of ourselves, assume that 
the template TF relation is known, and analyze residuals and measurement 
errors about that relation. We do so for the bivariate fit (see eqn. 18) 
to the sample displayed in Figures 21(b) and 21(c), the {\bf in+} composite
which includes all cluster members and peripheral objects as discussed
in section 6.3. The magnitude offsets produced by peculiar velocities
of individual clusters are removed before residuals are computed. The 
magnitude residual for the $i$--th galaxy is defined as $r_i = y_i - (a + b~x_i)$
where $(a,b)$ is the ``bivariate'' TF relation
of slope $b=\slop$ and offset $a=-21.01$.

Obtaining a reliable TF fit requires bias corrections which in turn
require a knowledge of the characteristics of the scatter. The empirical
determination of the latter requires that a reliable TF relation be known.
The circularity of the process forces an iterative approach, which
fortunately converges rapidly to a stable solution.

\subsection {The Total Observed Scatter}
 
In Figure 1 we summarize our estimates of the error budget in the TF diagram. 
The mean absolute deviation and the standard deviation of the observed 
residuals, averaged within bins in velocity width, are plotted respectively
as unfilled squares and circles. The observed values do
however underestimate the scatter: as discussed in section 3.2, sample
incompleteness depresses the measured values of the scatter at low velocity
widths. A correction for the incompleteness bias is applied to the observed
estimates of the scatter, shifting the points up: the bias corrected values
of the scatter are represented by filled circles and crossed squares. The 
locations of bias--corrected and measured scatter amplitudes
coincide for larger values of the velocity width. Indicative uncertainties for 
each bin are plotted as error bars about the corrected standard deviations; 
similar uncertainties apply to the values of the mean absolute deviations.

Note the decreasing trend in the scatter, with increasing velocity width. 
A linear fit to the scatter represented by the solid symbols is:

$$\epsilon_{tot} = -0.325 x + 0.32 ~~{\rm mag} \eqno(3)$$

\noindent
Deviations from the TF relation arise from three main causes:
\par \noindent
(i) measurement errors affecting the estimates of raw magnitudes, inclinations
and velocity widths;
\par \noindent
(ii) uncertainties arising from the corrections applied to measured parameters,
such as those for extinction, turbulent motions, etc.;
\par \noindent
(iii) ``cosmic'' scatter, an umbrella term for everything else, including effects
as varied as velocity field distortions, deviations from disk planarity, other
gravitational and photometric asymmetries, variations in the stellar
population make--up, variations in disk--to--bulge ratios, {\bf in}
cluster samples' depth, etc.
\par \noindent
Uncertainties arising from sources (i) and (ii) are estimated as described in
sections 4, 5 and 6 of paper VI; we refer to them as $\epsilon_x$ and 
$\epsilon_y$, where 
$\epsilon_x = 0.434\epsilon_w/W$ and $\epsilon_w$ is the error on the 
velocity width as given in eqns. (6) and (8) of paper VI. 
$\epsilon_x$ and $\epsilon_y$ are listed in Table 2 of paper VI, for 
each galaxy in the SCI sample.

\subsection {Measurement and Correction Errors}

What is the impact of measurement and correction errors on the TF scatter?
Given a generic function of two variables $f(x,y)$, the standard error
propagation formula indicates that

$$\epsilon_f^2 = \epsilon_x^2 \Big({\partial f \over \partial x}\Big)^2 +
\epsilon_y^2 \Big({\partial f \over \partial y}\Big)^2 + 
2 \epsilon_{xy} \Big({\partial f \over \partial x}
\Big)\Big({\partial f \over \partial y}\Big) \eqno(4)$$

\noindent In most cases, errors in $x$ and in $y$ are uncorrelated and the
covariant term $\epsilon_{xy}$ is assumed to be null. In the case of the
TF variables, however, errors in the two coordinates are correlated, 
mainly via the inclination dependences which correct velocity widths to
the edge--on perspective and magnitudes to a face--on aspect by means of an 
internal extinction term. It is thus more convenient to compute the scatter 
on the distance modulus inferred from the TF correction by expressing it in
terms of all the directly measurable quantities separately, so that
error contributions appear expressed in ways that can be easily compared.
The TF parameters $(x,y)$ in the relation

$$ y = a + bx \eqno (5)$$

\noindent
are obtained via the following correction recipes (see paper VI):

$$x = \log \Big[\Big({W_{obs} -\Delta_s\over 1+z} - \Delta_t\Big)/\sin i - 2.5\Big] \eqno(6)$$

$$y = m_{obs} - A_I - \Delta m_i + k_I - 5\log {cz-V_{pec}\over 100h} - 25 \eqno(7)$$ 

\noindent
where $m_{obs}$ and $W_{obs}$ are the observed magnitude and width; $\Delta_s$
is the width correction that accounts for the effects of instrumental broadening,
signal--to--noise distortions and data processing smoothing; $\Delta_t$ is the 
correction for turbulent motions; $i$ is the disk inclination; $A_I$ is the 
extinction arising in our
Galaxy; $\Delta m_i$ is the internal extinction term; $k_I$ is the cosmological correction
that accounts for the shift of the spectral energy distribution of the galaxy;
$V_{pec}$ is the galaxy's peculiar velocity, $z$ its redshift and $c$ the 
speed of light (see paper VI for details on each of the correction terms). 
The internal extinction term is defined in paper VI as 

$$\Delta m_i = -\gamma \log (1 - e) \eqno (8)$$

\noindent where $e$ is the ellipticity, corrected for seeing, and $\gamma$ is a
function of galaxy luminosity (or velocity width). The inclination is
derived from the ellipticity via

$$\cos^2 i = {(1-e)^2 - q_\circ^2 \over 1 - q_\circ^2} \eqno(9)$$

\noindent where $q_\circ$ is the intrinsic axial ratio of the spiral disk, for
which in paper VI we have assumed different values according to galaxy type.

Neglecting the contributions to errors arising from uncertainties on the 
redshift $z$ (typically less than 1\%), the extinction within the Milky Way 
(typically less than 0.02 mag), the $k_I$ term (typically less than 0.01 mag), 
and assuming a null peculiar velocity field, the error contributions on the
various observed and assumed quantities propagate to the distance modulus
via:

$$
\epsilon_m^2 = \epsilon_{mag}^2 + d_\gamma^2 \epsilon_\gamma^2 +  
d_{W}^2 \epsilon_{W}^2 +  
d_q^2 \epsilon_q^2 +  
d_e^2 \sigma_e^2,
\eqno (10)
$$
\noindent where

$$d_\gamma  = \log (1-e)$$
$$d_{W}     = 0.434 ~b/ W$$
$$d_q       = 0.434 ~b q_\circ /(1 - q_\circ^2)$$
$$d_e       = {0.434\gamma \over 1 - e} + {0.434 ~b (1 - e) \over 1 - (1-e)^2}$$

\noindent
Above, we have substituted (see section 5, Paper VI):

$$W = \Bigl[(W_{obs} -\Delta_s)^2/(1+z)^2 - \Delta_t^2\Bigr]^{1/2},$$

\noindent
which is the velocity width corrected for all but the disk inclination, so that

$$\epsilon_W^2 = {\epsilon_{W,obs}^2 + \epsilon_s^2 \over (1+z)^2} + \epsilon_t^2$$

\noindent where $\epsilon_{W,obs}$ is the total measurement error on the width.
For the remaining terms, we assume the uncertainties: 

$$\epsilon_q      = 0.15 q_\circ ,$$  
$$\epsilon_\gamma = 0.15\gamma ,$$ 
$$\epsilon_s      = 0.25 \Delta_s,$$
$$\epsilon_t      = 0.25 \Delta_t.$$

In Figure 2, we illustrate the relative contributions of the various components
to the total measurement error $\epsilon_m$, in two different cases. Curves
in Figure 2 have been calculated using the mean trend indicated by eqn. (3) of
paper VI, for the ellipticity; an intrinsic axial ratio for disks of 
$q_\circ = 0.13$; a TF slope $b=-7.67$; a measurement error on the magnitude 
of $\epsilon_{mag} = 0.05$ mag and the indicated values of $\gamma$. The 
curves labelled $\gamma, W, q_\circ$ and $e$ correspond to the terms 
$\vert d_\gamma \epsilon_\gamma \vert$,  
$\vert d_W \epsilon_W \vert$, 
$\vert d_q\epsilon_q \vert$ and 
$\vert d_e \epsilon_e \vert$.
In panel (a), which might correspond to a 
relatively bright galaxy ($\gamma$ is large) with well determined width 
($\epsilon_W/W = 0.02$), contributions to $\epsilon_m$  due to width, 
magnitude and ellipticity measurements are roughly equivalent; at high 
inclinations, the uncertainty on the value of $\gamma$ becomes dominant. 
On the other hand, panel (b) might correspond to the case of a narrow width,
low luminosity object (low value of $\gamma$), for which the relative error
on the width ($\epsilon_W/W = 0.07$) becomes the dominant source of 
$\epsilon_m$. In the latter case, the total measurement error is also 
substantially larger than in the first. This simulation helps illustrate 
important characteristics of the TF error budget, some of which are often 
misunderstood. First, except in exceptional cases errors in the photometry 
are generally negligible in comparison to those arising from other sources, 
such as the measurement and correction of velocity widths and those produced 
by internal extinction corrections. Second, velocity width measurement and 
correction errors are frequently, {\it but not always} the most important 
contribution to the total measurement error. Corrections
for internal extinction can dominate the total measurement error, e.g. for 
highly inclined, fast rotators.

\subsection {The Intrinsic Scatter Contribution}

For galaxies in our samples, we obtain averages of $\epsilon_m$ within bins of 
velocity width, shown connected by a solid line, alongside with the 
measurements of the total scatter in Figure 1. Since measurement and correction errors 
are independent of whether galaxies are located in clusters or in the field, mean
values of $\epsilon_m$ are computed from the joint sample of cluster and field 
galaxies (SCI+SFI), which allows a significant improvement of the statistical 
reliability of the estimates. We also plot, as dotted lines, the average
errors $|b|\epsilon_x$ and $\epsilon_y$, as defined in paper VI, which
represent the total errors on the two variables $x$ and $y$. The sum in
quadrature of $|b|\epsilon_x$ and $\epsilon_y$ approximates but does not match
exactly the value of $\epsilon_m$, because of the coupling of errors through
the inclination correction; on the average, such coupling is mild.
 
While they constitute an important fraction of the total budget, 
measurement errors alone cannot fully account for the total amplitude of 
the scatter about the TF relation. In Figure 1, we plot as dashed lines
the combinations in quadrature of the total measurement error with  
components of ``intrinsic'' scatter $\epsilon_{int} = 0.20$, 0.25 and 
0.30 mag, respectively from bottom to top. While the trend in the observed 
residuals is qualitatively reproduced by these combinations, the slope that
results from the assumption of a constant value for the intrinsic scatter 
does not adequately track the data. It appears instead that while a value
of $\epsilon_{int} \simeq 0.20$ might match the residuals at the high
width end, a value as high as $\epsilon_{int} \simeq 0.35$ might be necessary
for low velocity widths. The trend in $\epsilon_{int}$ that best fits the data
is:

$$\epsilon_{int} = -0.28 x + 0.26 \eqno(11)$$

\noindent
The combination of eqn. (11) and the smoothed trend of $\epsilon_m$ is plotted
as a solid, heavy line, which matches well the observed trend in the standard 
deviation of the residuals (filled circles). Because the distribution of residuals 
about  the TF relation exhibits a slight measure of kurtosis, caused by physical 
and/or spurious outliers, it could be argued that our error model 
should track somewhere below the standard deviations, and somewhat closer to the
mean absolute deviations (crossed squares). We prefer to adopt the conservative 
approach indicated by eqn. (11), however, with the proviso that the intrinsic
component of the TF relation scatter is unlikely to exceed the values predicted
by eqn. (11).

A possible contributor to the ``intrinsic scatter'' component is
the spread in distances associated with galaxies in each {\bf in}
cluster sample. Since all galaxies in such a sample are not
at the same distance, their spread in distance results in a scatter
in magnitudes roughly in the measure $\epsilon_{dist} \sim 0.10 
\theta_{rms}$ mag, where $\theta_{rms}$ is a measure of scatter in 
the sky distribution of galaxies. The average contribution for the
clusters in our sample is between 0.10 and 0.15 mag, although the TF 
scatter for nearby, dispersed groups, such as Eridanus, may be 
dominated by their physical dispersion. We return to this issue
in section 5 and in Figures 19 and 20.

TF fits that yield r.m.s. scatter figures smaller than 0.25 mag are likely 
to be statistical accidents, which can occur when samples are small (e.g. 
Bernstein \etal 1994 for Coma, Mathewson \etal 1992 for Fornax).

\subsection {Some Implications of the TF Scatter Behavior} 

Franx and de Zeeuw (1992) have found that the TF scatter poses strong 
constraints on the elongation of the gravitational potential in the
disk plane of spirals. Our estimate for the ``intrinsic'' part of the TF 
scatter reinforces the conclusion of Franx and 
de Zeeuw, that the average ellipticity of the potential in the plane
of the disk must be smaller than about 0.06. On the basis of 2.2 $\mu$m
photometry of 18 face--on spirals, Rix and Zaritsky (1995) find that
departures from disk axisymmetry may contribute $\sim 0.15$ mag to the 
TF scatter. Their conclusions apply principally to the inner
regions of the disk, which are sampled by the K--band observations, and
therefore to TF scatter based on optical observations of the rotation
field. When the TF relation is principally based on 21 cm rotational
data, the effect of disk asymmetry on the scatter becomes more ambiguous,
as it relates to outer and more poorly mapped regions of disks. Eisenstein
and Loeb (1996) estimate that the scatter resulting from varying formation
histories of galaxies should exceed 0.3 mag for a broad class of cosmological 
scenarios. The relatively low values inferred for $\epsilon_{int}$ suggest 
either an unexpectedly late epoch of galaxy formation or that a form of 
secular, regularizing feedback mechanism may be responsible for the 
tightness of the TF relation, as suggested by Silk (1996).

Sandage and collaborators (1994, 1995) have advocated a large value of the
TF scatter, near or larger than 0.7 mag, as an explanation for the high
values of the Hubble constant resulting from the use of the TF relation. 
If the scatter were as large as proposed by that group, large biases
would result, the adequate correction of which would lead to a lowering
of the zero--point of the TF template relation, and a consequent reduction
of the value of $H_\circ$. While the values of the scatter found by us are
not as low as advocated by other groups, it appears unlikely that in the
I band the scatter may be as large as suggested by Sandage \etal .

It has been advocated that the use of an {\it inverse} fit if the TF relation 
--- one where the ``independent'' variable is the magnitude rather than the
velocity width --- does away with the need to correct for incompleteness
bias (e.g. Schechter 1980). The nature of the TF scatter, 
especially the fact that velocity width errors can be overshadowed by
other sources, weakens the case for a bias--less inverse TF relation.
We return to this issue in section 3.2.

\section{Bias, Morphological and Environmental Corrections}

It has been argued that many of the thorny problems associated with the
derivation of a TF relation that arise with field samples are avoided
when cluster samples are used. While cluster samples do indeed offer many
advantages, they are affected with biases that can severely influence the
inferred TF relation. While the optimists are well justified in promoting
the adoption of cluster templates over field ones, so are the pessimists in
advising skepticism regarding extravagant claims of the curative properties 
of the cluster elixir. Following seminal work by Roberts, Bottinelli, 
Gouguenheim, Paturel and Teerikorpi,
the discussion of bias in cluster samples' TF has been extensively discussed
in the literature, and important points have been underscored in the analyses
of Bottinelli et al. (1986), Teerikorpi (1984, 1990, 1993), Kraan--Korteweg et 
al. (1988), Fouqu\' e et al. (1990), Pierce and Tully (1992), Willick (1994) and
Sandage \etal (1995).

Suppose a universal TF relation (UTF) exists. An observed sample can depart
from the UTF for a variety of reasons:

\noindent
(a) If a cluster sample consists of galaxies of various morphologies: 
do galaxies of all types abide by the same UTF? We consider this question
in section 3.1.

\noindent
(b) An important source of bias results from the coupling of a cluster sample 
completeness with the scatter in the TF relation. This effect has been 
recently illustrated in great detail by Sandage and collaborators (Sandage 
1994a; Sandage 1994b; Federspiel et al. 1994; Sandage \etal 1995); their 
graphic approach allows a rapid conceptual grasp of the problem. We discuss 
this problem and our adopted solution in section 3.2.

\noindent
(c) Samples extracted from size-- or flux--limited catalogs, such as the UGC 
(Nilson 1973) or the CGCG (Zwicky \etal 1961), will in varying degree have 
built--in biases, which depend on the accuracy of the cataloged photometric 
quantities which were used to select the sample, rather than on that of the 
successively measured parameters. In particular, biases can arise from the 
proximity of sample objects to the catalog limit, and they may be strongly 
coupled to the amplitude of the scatter in the TF relation. This ``edge of 
catalog'' effect can be folded in with the selection function of the sample; 
in the treatment advocated by Willick (1994), the correction includes that of 
the incompleteness bias we have discussed above, and becomes the dominant 
bias correction. We discuss it separately in section 3.3.

\noindent
(d) When the size of the cluster is not negligible in comparison to its
distance, the estimated mean distance to a cluster can be in error. The 
effect, which is usually quite small, is briefly discussed in section 3.4.

\noindent
(e) The so called ``homogeneous Malmquist bias'' is important in the correction 
of the estimated TF distances of individual field galaxies, and is computed 
on the assumption of a Poissonian distribution of galaxies in space. This 
bias, which couples with the characteristics of the scatter in the TF diagram,
is much smaller in the case of clusters than for individual field galaxies.
In section 3.5, we derive an estimate for such bias, taking into account
the variable character of the scatter in the TF plane, and discuss its
implications for the determination of cluster distances.

\noindent
(f) The cluster environment may affect both the photometric and kinematical
characteristics of galaxies, depending on their location in the cluster
gravitational potential well. As a result, the location of individual objects 
in the TF diagram may depend on distance from the cluster center or on the
cluster richness. We explore this possibility in section 3.6.

\subsection{TF Dependence on Morphological Type}

The work of Roberts (1978), de Vaucouleurs et al. (1982) and Burstein (1982) 
raised a serious concern 
on the predictive quality of the TF technique, by suggesting that a single
relation may not apply to all morphological types, a concern expressed
by Tully and Fisher (1977) themselves. While it appeared
clear that different types follow different TF relations when B band 
photometry is used, the work of Aaronson and Mould (1983)
demonstrated that such dependence becomes negligible when H--band infrared
magnitudes are adopted. The question resurfaces in this analysis, for it
has not been satisfactorily verified whether a single TF relation is
applicable to all types at I band.

The issue is important for us, because the SCI, as any TF cluster sample,
spans a broad range of spiral types. This circumstance arises because of the 
relative paucity of spirals in clusters, as compared with the field. In the 
field it is feasible to restrict samples to contain only late spiral types 
(i.e. Sbc and Sc). In clusters, however, the scarcity of spirals induces 
observers to stretch somewhat the morphological type constraints, in order
to accrue statitically meaningful samples. In the overall {\bf in+} sample, 
for example, 7\% of the galaxies are typed Sab or earlier, 30\% are Sb's, 
57\% are Sbc--Sc's and 6\% are of morphology later than Sc. 

Consider Figure 3(a), where we plot with different symbols the location
in the TF diagram of SCI galaxies earlier than Sb (large
solid circles), Sb's (dotted squares) and later than Sb (small dots).
Differences in the behavior of the three classes appear. 
No sample incompleteness correction (discussed in
the next section) has been applied to the data in Figure 3. A fiducial (dashed)
TF line of slope \slop is superimposed on the data, and it represents the
best fit to the types later than Sb. A mild, perhaps monotonically 
changing dependence of the offset of the TF relation on type is seen; earlier 
types appear to be slightly fainter, at a given width, than later ones.

We have also investigated possible differences in the behavior of the galaxies
later than Sc. These objects are, however, few in number in our cluster
sample, and their scatter 
about the TF relation is typically large. They appear to be somewhat brighter,
at a fixed width, than the other types. However the indication among similar
objects in the field points in the opposite direction, i.e. they appear to
be somewhat fainter than the other types, at fixed width. Given the unclear
picture, we thus choose not to fit them separately; in Figure 3, they are 
included in the late spirals group.

It is well known that galaxies of different type have different distributions 
of velocity width (Roberts 1978): earlier types are typically faster rotators
than later types. The differences suggested by Figure 3(a) could then result
from both luminosity shifts or changes in TF slope. In Figure 4(a),
we plot solely the galaxies earlier than Sb, combining  objects
in the cluster sample (SCI; solid symbols) with those in the field sample (SFI),
in order to help improve statistical significance.
In Figure 4(b), the cluster Sb galaxies are shown. In each of the
panels of Figure 4, the dashed line represents the fiducial TF line of slope
\slop also plotted in Figure 3. 

Linear fits to the data in Figure 4 indeed yield shallower slopes than for the 
later type spirals. However, this impression is tempered by the fact that
the early--type spirals cover a narrower range of luminosities and are affected 
by an incompleteness bias of the type decribed in the next section. 
The bias produces an apparent reduction of the TF slope, and it may well be 
stronger for the early type galaxies than for the later types. When the effect 
of incompleteness is taken into account, the difference in slope between samples 
of different type becomes statistically insignificant. We thus adopt a 
morphological type correction which is a plain additive offset:

{\it Types earlier than Sb: $-0.32$ mag}

{\it Type Sb: $-0.10$ mag}

{\it Other types: unchanged}

Figure 3(b) displays the same objects as Figure 3(a), except that galaxies of types 
earlier than Sbc are displaced by the morphological type correction described above.

The type correction adopted for Sb galaxies is only marginally significant. 
Nonetheless, the 0.10 mag offset is corroborated by inspection of the field
galaxy sample; it is also in agreement with independent estimates
of mass--to--light ratios (Broeils 1992; Vogt 1995). 

Early--type spirals occur preferentially in richer clusters, which 
tend to be more distant in our sample. Since the type correction is more 
conspicuous for the early types, the concern arises that the type 
correction may mask a distance bias. We can discount such concern.
The type bias is corroborated by the behavior of galaxies in the field sample, 
as illustrated in Figure 4(a), and by direct comparison of objects of the 
same average distance but different type.

We close this section with a few comments regarding the issue of
linearity of the TF relation. It is interesting to point out that the global 
TF relation shown in figure 3(a), before any type bias is accounted for, 
exhibits a mild degree of nonlinearity of the type noted by several authors
in the past (i.e. Aaronson et al. 1982; Willick 1990; Pierce and Tully 1996). 
We have discussed
in paper II (see Figure 9 in that ref.) how such nonlinearity can arise in 
part from the adoption of an inadequate extinction relation. An additional 
contribution to nonlinearity can result if samples contain a mixture of 
morphological types, and their differential behavior is ignored. 
There are obvious advantages of numerical simplicity in the adoption of a
linear model. However, it is important to point out that there are no
well established physical reasons which demand strict linearity in the
TF relation (see Giovanelli 1996).

\subsection{Incompleteness Bias}

Echoing a remark of
Sandage \etal (1995), we stress the point that a cluster sample, just
because all the galaxies are at roughly the same distance, {\it does not
exhibit the properties of a volume limited sample}. The parameters of a 
cluster TF diagram will depend on (a) the degree of completeness with which 
the luminosity function (LF) is sampled and (b) the amplitude of the scatter in
the unbiased UTF. Roberts (1978) first discussed the effect of
a Malmquist--like bias on the TF relation, particularly in connection with
its application in measuring the value of $H_\circ$. Schechter (1980) pointed
out that the adoption of an {\it inverse} fit, where absolute magnitude is
used as the `independent variable', yields bias--free predictions of distance.
An analogous approach has been adopted by several workers (e.g. Pierce and Tully 
1992) and discussed extensively (see Teerikorpi 1990, 1993 and references therein).
Conversely, the derivation of a large bias correction, based on the assumption
of a TF scatter of large amplitude, has been used by Sandage et al. (1995) as a 
means to explain why previous  use of the TF relation may have led to 
overestimates of the value of $H_\circ$.

If samples were selected purely by angular size or flux criteria, and 
subsequent extinction corrections were negligible, and if furthermore all
observational and other correction errors had negligible impact on magnitudes,
then the inverse TF relation would indeed yield nearly bias--free predictions
(see, e.g. simulations in paper III). However, since catalogs are inexact,
extinction corrections can be large and thus shift points in magnitude,
and since errors in corrected absolute magnitude can not only be important
but in some circumstances even predominant (e.g. Figure 2a), the correction
for bias becomes an unavoidable chore, whichever fitting convention is
adopted. In the following, we describe the nature of the bias, our adopted
correction technique and its sensitivity to various relevant parameters.

\subsubsection{The Nature of the Bias}

The nature of the incompleteness bias is dramatized in the simulation
shown in Figure 5. A cluster sample is extracted from a population with
an LF as shown by the smooth curve plotted along the 
vertical axis; the sample is however incomplete, as the histogram of
magnitudes superimposed on the LF shows. Incompleteness
is shown here to have a ``soft'' edge, indicated by the progressive
departure of the histogram from the LF. For a flux limited
sample, the histogram would of course track the LF for
magnitudes brighter than the limit, then level suddenly to zero. The assumed 
UTF is represented by a dashed line. In order to dramatize the bias, we
have assumed a scatter twice as large as that illustrated by eqn. (3);
the assumed scatter in Figure 5 averages about 0.7 mag. A heavy solid 
line connects filled symbols which identify mean values of the magnitude
within bins of velocity width. Incompleteness affects the TF relation 
derived from the simulated sample in several important ways:

\noindent
(1) the derived slope is {\it less steep} than that of the UTF; the
slope becomes progressively shallower as the velocity width diminishes
because, due to the incompleteness, low width objects preferentially
appear above the UTF; 

\noindent
(2) the zero point is {\it brighter} than that of the UTF; 

\noindent
(3) the scatter is {\it underestimated}, increasingly so as the velocity
width diminishes, because objects with high residuals below the UTF line
are missing from the sample.

\noindent
As a result of this effect, one should expect that uncorrected cluster TF 
slopes will diminish with increasing distance. This expectation needs to
be tempered by noting that the amplitude of the scatter, and thus of the bias,
is actually smaller in reality than indicated in the simulation of Figure 5.
In spite of the broad variance introduced by the small number of objects per 
cluster, inspection of Figure 17, which contains TF diagrams of the SCI 
clusters uncorrected for the incompleteness bias, shows that the steeper 
TF slopes correspond to Ursa Major and Fornax, two nearby clusters which 
sample fairly deeply the LF; on the other hand, the distant
clusters Coma and A2634, represented by samples which become severely 
incomplete near M$_I + 5\log h = -20$, have among the shallowest TF slopes.
One should also expect the uncorrected TF zero point to become progressively
brighter with increasing distance. This effect is small and not discernable
in the figures, where zero point variations are dominated by cluster motions. 
A change in the scatter of 
individual cluster TF diagrams with distance is also difficult to discern 
from the plots in Figure 17, due to the smallness of the effect and of the 
size of individual cluster samples.

In this section we have illustrated the nature of the bias by reasoning
within the framework of a {\it direct} TF relation. Since errors are important
in both coordinates, a description of bias could analogously be produced
for the inverse relation. As we shall discuss in section 4, our favorite
fitting mode will be one that takes however into consideration errors in 
both coordinates.

\subsubsection{Our Correction Technique. Case study: the Coma Cluster}

Having established the nature of the incompleteness bias, we must decide 
how to correct for it. Several papers have given correction recipes, mostly
attempting to produce analytical or graphic solutions to the problem 
(Teerikorpi 1993 and references therein; Willick 1994; Sandage \etal 1995).
These treatments usually assume a single value for the TF scatter, a sharp
apparent magnitude or angular size limit for the completeness of the sample,
or both. These conditions are not met by realistic samples. We have
emphasized in section 2 that the TF scatter varies significantly with velocity
width, mostly driven by measurement errors associated with the velocity widths, 
disk inclination and correction uncertainties associated with turbulent motions
and internal extinction. An inspection of the
completeness histograms plotted for each cluster in Figure 10 shows not only the 
varying degree, but also the
softness of the limit of completeness for each cluster sample. Under these 
circumstances, the incompleteness bias problem is more readily solvable 
numerically, adopting Monte Carlo techniques. In what follows, we illustrate 
our correction method, step by step, for the case of the Coma cluster.

Panel (a) of Figure 6 displays the ``raw'' TF diagram for the {\bf in+} sample
of Coma. The small unfilled symbols represent the positions of the galaxies as
listed in Table 2 of Paper VI. A type correction is applied to galaxies of type
earlier than Sbc. These are typically fast rotators. The type--corrected 
positions are shown as filled symbols and the direct fit to the type--corrected 
data points as a dashed line. In the same panel, plotted along the 
left vertical axis is a shaded histogram of the absolute magnitudes of the 
sample and the LF of the spiral population, which we use 
for the estimate of sample completeness. Incompleteness sets in near $-21$ 
and sampling of the LF ceases below $-20$. The estimate of 
the incompleteness bias requires several ingredients: 

\noindent (i) a function which describes the sample completeness, $c(y)$;  

\noindent (ii) a UTF of zero point and slope 
$(a_{utf}, b_{utf})$; 

\noindent (iii) a description of the unbiased TF scatter 
characteristics. 

\noindent For each cluster sample, the completeness function is
constructed as follows. First we Hanning--smooth (i.e. convolve with a 
[0.25,0.50,0.25] function) the magnitude histogram, in order to partly
reduce the effects of small number statistics. The smoothed histogram
is then divided by the LF, producing a histogram $h'(y)$. 
The ratio $h'$ is then fit with a function

$$c(y) = {1 \over e^{(y - y_F)/\eta} + 1} \eqno(12)$$

\noindent
(which is the shape of the energy distribution function for the Fermi--Dirac 
statistics and provides a familiar example of a smooth step function). 
The parameter $y_F$ is the absolute magnitude 
at which sample completeness approximates 50\% and $\eta$ parametrizes
the steepness of the completeness decline. We use the analytical fit
to describe completeness, rather than the histogram $h'$, in order to
reduce the vagaries of small number statistics. Both the histogram
$h'$ and the function $c(y)$ (dashed line) for Coma are shown in the
appropriate panel of Figure 10. 

Operationally, the estimate of the incompleteness bias is 
carried out by producing $n_{times}$ simulated realizations of the cluster. 
Each realization has the same distribution of line widths as the real sample. 
A simulated sample point has a given velocity width parameter $x$ (extracted 
from the real cluster sample); from it, a value $y = a_{utf} + b_{utf}x + 
\epsilon (x)$ is computed, where $\epsilon (x)$ is the function described by 
eqn. (3); the point is then passed through a ``completeness filter'' by 
comparing the value of the completeness function $c(y)$ for that magnitude 
with a random number $n_{ran}$ in the interval (0,1): if $n_{ran} < c(y)$ 
the point is accepted, and if the condition is not met the point is rejected 
and a new value of $y$ is generated and tested. The simulated sample will 
reflect the incompleteness characteristics of the real sample; its biases, 
which can be well estimated if $n_{times}$ is large, are the most likely 
estimate of bias for the real sample. The simulated sample is fit by a 
linear relation $(a_{fit},b_{fit})$. 

Panel (b) of Figure 6 shows the mean values of the deviation between the 
simulated sample points and the UTF relation, after binning the data in 
narrow velocity width intervals and $n_{times}=800$ trials. This is the 
incompleteness bias for the Coma cluster.

Panel (c) of Figure 6 shows the average values of the ratio of the r.m.s. 
scatter computed for the simulated sample, with respect to the TF relation 
$(a_{fit},b_{fit})$, over the ``true'' value input for the UTF. We appreciate 
here how the simulated sample underestimates the true scatter of the TF 
relation at the small width end. 

Figure 6(d) is an approximation to an unbiased representation of Coma's TF 
diagram. To each point in figure 6(a) a correction in $y$ is applied to 
account for the bias shown in panel (b), i.e. each point is shifted 
faintwards. The sample is still incomplete, and the scatter at the lower 
widths is slightly underestimated, but the UTF slope and zero point are
recovered free of bias. The dashed line is the same as in panel (a).

Next we investigate the sensitivity of the bias estimate on our input 
parameters. In Figure 7 we display panels analogous to those in Figure 
6, panels (b) and (c), 
in which the bias is plotted for three different functions describing the 
scatter $\epsilon (x)$. The solid line shows the bias for our adopted 
function $\epsilon (x)$, as given in eqn. (3), while the other two lines 
illustrate the bias if the scatter were respectively 1.5 times and twice 
as large as the adopted amplitude. In Figure 8, 
we maintain the scatter function $\epsilon (x)$ fixed at our adopted form, 
and alter the slope of the faint end of the LF, which 
affects the completeness function of the sample. Using a Schechter form 
for the LF, the two cases plotted correspond to exponents
of $\alpha=-0.50$ and $\alpha=-0.85$, respectively, for the power law,
faint--end part of the functional representation (the value of $\alpha$
used in Figures 6 and 7 is $-0.50$). The two functions are shown as an inset in
Figure 8b. Finally, in Figure 9, we maintain scatter function $\epsilon (x)$ and
LF ($\alpha=-0.50$) fixed, and change the slope of the UTF between values 
of $-7.0$ and $-8.0$. {\it It is clear that the main driver of the 
bias amplitude is the TF scatter}. Changes of the UTF slope affect only in 
minor form the estimate of bias. The slope of the LF, while
not affecting in great measure the overall amplitude of the bias, shifts
its effects to galaxies of higher velocity width (brighter); as a result,
it has an effect on the zero point of the TF relation: in the case of
Coma the zero point shift, resulting from a change between the two 
LF's displayed in Figure 9, amounts to about 0.04 mag.

\subsubsection{Incompleteness Bias Corrections for our Cluster Samples}

In Figures 10 and 11, we summarize the incompleteness bias computations for 
each of the {\bf in+} cluster samples in SCI. They were computed using a
UTF slope of -7.6, a LF index $\alpha = -0.5$ and
a functional description of the scatter amplitude as given by eqn. (3).
The coefficients $y_F$ and $\eta$ of the completeness function as defined 
by eqn. (12), computed for each cluster {\bf in+} sample, are listed in cols. 
(2) and (3) of Table 1.
Each bias estimate was obtained after $n_{times} = 800$ simulations of
the cluster.

In Figure 10, the shaded histograms represent the functions $h'$, obtained
as discussed in the preceding section. The superimposed dashed lines 
represent fits of the type described by eqn. (12). The completeness level
of $c(y)=1$ has been estimated by eye as shown, compatibly with the noise of
the $h'$ histograms. The parameters $y_F$ and $\eta$ are not computed 
separately for the {\bf in} samples. Because the {\bf in} samples contain 
typically half the number of objects of the {\bf in+} samples, their $c(y)$ 
functions would be quite unreliable. However, for any given cluster 
the luminosity distributions of {\bf in+} and {\bf in} samples are very similar,
and it can be safely assumed that the $c(y)$ functions derived for 
{\bf in+} samples are applicable to the {\bf in} samples as well.
The clusters A2197 and A2199, which are close in the sky and at approximately
at the same redshift, are combined in a single sample in Figure 10.

In figure 11, we display the results of our computations for the 
incompleteness bias for each cluster sample, obtained following the
procedure described in detail in the preceding section for the Coma
cluster. A solid line superimposed on the results of the simulations
indicates the adopted bias correction. Notice how incompleteness affects
the TF diagrams of the more distant clusters, such as A2634 and A2197/9,
over the whole dynamic range covered by the diagram.

We have also computed the analogous functions to those displayed in Figures
10 and 11, using an exponent $\alpha = -0.85$ for the LF. The corresponding 
coefficients $y_F$ and $\eta$ of the completeness function as defined by eqn. 
(12) are listed in cols. (4) and (5) of Table 1. We do not show the graphic
results (analogous to figs. 10 and 11) for the sake of economy. The changes
with respect to the functions displayed in Figure 11 can be garnered by 
inspection of Figure 8, for Coma. The variance in bias estimates which results
from different assumptions of the LF shape will be used in section 6, to 
investigate the margin of uncertainty on the determination of cluster motion 
offsets and on the zero point of the TF template relation.

In Figure 12, we plot the bias corrections for each galaxy in the {\bf in+}
samples, as a function of the I--band apparent magnitude. In panel (a), the
correction has been computed using a value of $\alpha = -0.5$ for the power
law exponent of the LF, while in panel (b) the value
$\alpha = -0.85$ was used.

We remind the reader that the bias corrections illustrated here were
estimated within the framework of the bivariate TF relation, where
a direct fit takes into consideration errors in both the variables,
as described in section 4.

\subsection{The Edge--of--Catalog Bias}

We start with a simple example, which illustrates the concern addressed 
in this section.
Consider the ideal case where a flux--limited sample for a given cluster is 
extracted from a catalog, so that the sample can be considered complete for
all magnitudes $m$ up to a limiting value $m_l$. Note that even if the selection 
were strict, the sample would never be genuinely complete, because the 
catalog itself will have built--in biases, which can be serious, 
as discussed extensively by the Cardiff group: see Davies 1994 and references 
therein. A TF relation with given scatter is used to infer ``predicted
magnitudes'', $m_{tf}$, from velocity widths of each galaxy in the sample.
The differences $m_{tf} - m$, when plotted versus $m_{tf}$, will exhibit a
positive bias for values near and larger than $m_l$: the scatter in the
TF relation shifts some objects to values of $m_{tf}$ larger than $m_l$.
As in such a sample the number of objects per magnitude bin, $n(m)dm$,
rises steeply up to $m_l$, the mean estimated distance averaged over
$m_{tf}$ bins is biased near $m_l$.
 
Willick (1994) has developed an elaborate mathematical formalism to treat
this effect, which can be extended to cases in which samples are not strictly 
magnitude-- or size--limited. For a given cluster sample, for example, he 
estimates the bias in terms of 5 parameters: the scatter
in the TF relation, two so--called ``coupling parameters'', which relate
to the sample selection functions (in magnitude and diameter), and two
``closeness parameters'', which are estimated for each galaxy from its
velocity width, the TF scatter and the size and magnitude limits of the
sample. The inclusion of the coupling parameters makes it possible for
this approach to treat jointly the edge--of--catalog effect and the 
incompleteness bias described in the preceding section. However, the
incongruous character of each sample's selection functions and the
inadequacy of the assumption, made in the derivation of Willick's
formulas, that the TF scatter is a constant 
(see discussion in section 2), raise complications in the application
of this technique.
We test for the severity of a catalog edge effect in our data in two ways.

First, we reproduce the correction estimated in Willick (1994, his Figure 1b)
for a strictly magnitude--limited sample,
by means of a Monte Carlo simulation and following procedures similar to
those described in section 3.2. We display the results in Figure 13, where
we use Willick's nomenclature to define the variables 
$A=(m_l - m_{tf})/\sqrt{2}\sigma$ and $B=-(m-m_{tf})/\sigma$, where
$\sigma$ is the scatter in the TF relation, which Willick adopted to be
constant. We carried out simulations adopting for $\sigma$ both a constant 
value and a variable one as described in eqn. (3); in the latter case,
its mean value over the simulated sample was adopted in the definition
of $A$ and $B$. The results of our simulation for a magnitude--limited 
sample are represented by
the unfilled squares in Figure 13, while Willick's analytical estimate
is shown as a continuous curve. When, instead of a sharp
magnitude edge, we apply to the simulated sample a completeness
function as estimated for our global {\bf in+} sample, the
expected bias, shown as solid circles, is indistinguishable from zero. 

The reason for a lack of an edge--of--catalog bias in our data is that
there is no hard edge in our samples at the cataloged magnitude limit;
incompleteness starts at significantly brighter magnitudes than either
the CGCG (Zwicky et al. 1961) magnitude limit or the Lauberts (1982)
limit, and it falls off gradually on the faint end. 
A corroboration of the lack of an edge--of--catalog bias can
be obtained by inspecting the TF residuals as a function of redshift, 
as shown in Figure 24. Since galaxies near the catalog 
edge tend to be more distant objects, a catalog edge effect would appear
in the form of systematically negative residuals, at high redshift. Such an
effect, or any trace of systematicity, are not present in Figure 24.

We thus choose not to apply to our data a correction for the catalog edge bias
which, if residually present after the incompleteness bias is applied, 
appears buried in the noise.

\subsection{Cluster Size}

The assumption that all galaxies in a cluster are at the same distance
from us is adequate insofar as the ratio between the cluster physical
size and its distance is negligibly small.
Such an assumption is increasingly flawed as the distance of the cluster
diminishes. There are two kinds of corrections that may need to be applied,
in order to account for the extent of clusters along the line of sight.
One relates simply to the mismatch which results from working with
distance moduli rather than actual distances; this correction is
applied to the mean distance of the
cluster ensemble. The second needs to be applied to each galaxy,
and differs depending on the nearness of the galaxy to the completeness
limit of the sample; this correction is related to the incompleteness
bias. We consider them separately.

\subsubsection{Cluster Size: Global Bias}

When we fit a cluster TF diagram, we minimize the scatter on the
distance modulus, not on the distance. We thus obtain a mean distance 
modulus which may differ from the mean distance of the cluster, especially 
if the ratio between diameter and distance  of the cluster sample is high. 
This difference arises because the average of the logarithm is not the 
logarithm of 
the average. We estimate this bias by assuming that the distribution
of galaxy distance in each sample is Gaussian, with a dispersion
which we estimate from the angular spread in the sky. The difference
between the mean and the true distance modulus can be expressed
in terms of the ratio between the size, expressed in terms of the
Gaussian dispersion of the galaxy distribution, $\delta$, and the distance
$d$ to a cluster: it amounts to -0.08, -0.06, -0.03, -0.008 mag
respectively for $\delta/d$ 0.20, 0.13, 0.10, 0.05 radians. For clusters 
in our sample, the effect is typically small. Estimates of the bias
for each sample, expressed in magnitudes, are given in cols. (6) and
(7) of Table 1, respectively for the {\bf in} and {\bf in+} samples
of each cluster.

\subsubsection{Cluster Size--Sample Incompleteness Bias}

Let the distribution of galaxy distances within a cluster be
Gaussian, with a dispersion which can be translated to $\sigma_d$
mag e.g., if the dispersion in distances along the line of sight
is 10\% of the mean distance, $\sigma_d = 0.21$ mag. When the assumption
is made that all galaxies in the cluster are at the same distance from us, 
the adopted absolute magnitude $y$ of a given galaxy will be off the true
value according to a Gaussian probability of dispersion $\sigma_d$.
If the sample is complete at magnitude $y$, the mean expectation value
of $y$ will be equal to the true value, i.e. the probability that the
galaxy is in the foreground half of the cluster is equal to that of
finding the galaxy in the background half. However, if the sample 
completeness decreases steeply near $y$, it will be more likely that
the galaxy be placed in the foreground half than in the background half of
the cluster. Assigning to the galaxy the mean distance modulus of the
cluster will lead to an overestimate of its luminosity.

The effect described is illustrated in Figure 14. The curve labelled
$c(y)$ is the completeness function of the sample, as parametrized by
eqn. (12); the curve labelled $E(y)$ describes the Gaussian extent
of the cluster, whereby galaxies of absolute magnitude $y_\circ$ are
seen spread with dispersion $\sigma_d$ (for the computation of this
bias we assume that a Gaussian distribution in size translates into a
Gaussian distribution in distance modulus; the marginal validity of this 
assumption affects the bias estimates negligibly, only in the second
order); the heavy--traced curved is the product $c(y)E(y)$. 
The bias that arises from the coupling of cluster size and sample 
incompleteness is the difference between the centroid of $c(y)E(y)$
and that of $E(y)$, i.e.

$$\beta_{EI}(y_\circ)= \bar y - y_\circ \eqno (13) $$

\noindent where

$$\bar y = \int y c(y) E(y) dy / \int c(y) E(y) dy$$
\noindent Quantitatively, the effect
is generally small. For example, in the case of a cluster with a
line of sight extent equal to 10\% of its distance, and a completeness
function of the type given by eqn. (12) with $\eta = 0.5$ mag, the
bias at $y_\circ \simeq M_F$ is $\beta_{EI}\simeq -0.05$ mag.

The correction $\beta_{EI}$ should in principle be applied only
to cluster galaxies, i.e. objects in the $\bf in$ samples, for which
the assumption of a common distance was made. Galaxies in the {\bf in+}
samples were assigned the distance derived from their individual 
redshift, and such a bias correction would appear at first not to be 
necessary. However, a measure of scatter about their velocities
should be expected for the motions of field galaxies as well. Such scatter 
is equivalent to describing the distance by a probability distribution,
centered about that corresponding to the redshift velocity, which will 
be present even 
in the absence of large scale velocity flows. A bias similar to
that described above for cluster galaxies thus arises, when the
absolute magnitude of the galaxy approaches values for which the
sample is incomplete. For the non cluster galaxies
pertaining to the {\bf in+} samples, we can thus compute a 
bias similar to $\beta_{EI}$, where the function $E(y)$ is
assumed to be a Gaussian of dispersion $\sigma_d = 5\log (1+\sigma_v/cz)$,
with $\sigma_v \simeq 200$ \kms .

\subsection{Homogeneous Malmquist Bias}

The so--called Malmquist bias produces statistical underestimates of 
TF distances. This arises from the fact that within a given solid angle,
the number of galaxies between distances $r$ and $r + dr$ usually rises 
with $r$. Then, for galaxies of estimated distance modulus 
$\mu_e \pm \Delta_\mu$, the most probable distance is not $r_e = 
10^{0.2(\mu_e-c)}$ (where $c$ is the usual scaling term which depends on 
the adopted units of distance), but a value $r>r_e$, because there are more 
galaxies between $\mu_e + \Delta_\mu$ than between $\mu_e - \Delta_\mu$. 
When the assumption is made that the space distribution of galaxies is 
Poissonian, the qualification of ``homogeneous Malmquist bias'' is
used. In that case,

$$r = r_e e^{ {3.5\Delta^2}} \eqno(14)$$

\noindent
where $\Delta = 10^{0.2\Delta_\mu} - 1$ is the relative error 
in the distance estimate arising from $\Delta_\mu$. For example, for a
scatter of $\Delta_\mu = 0.3$ mag, the distance relative error is 15\%,
or $\Delta = 0.15$. For recent reanalyses of this classical problem
(Malmquist 1924), relevant to the measure of galaxy distances, see 
Lynden--Bell \etal (1988), Feast (1987), Landy and Szalay (1992),
Sandage (1994), our Paper III and references therein.
The estimate of the bias in a clustering regime is very different 
from the homogeneous case and will depend on the characteristics of
the space density field.

To the extent that clusters can be characterized as single objects in 
space, a Malmquist bias correction (MBC) in principle also applies to their 
estimated distances. However, since the distance modulus determination of 
a cluster is more accurate than that of a single galaxy (roughly by a 
factor $N^{1/2}$, where $N$ is the number of galaxies measured in the
cluster), the amplitude of the MBC is quite small. 

Our analysis of the error budget of the TF relation shows that $\Delta_\mu$
depends on the galaxy velocity width, and thus so will the Malmquist bias.
In Figure 15 we show the homogeneous MBC, $r/r_e$, computed assuming that the
TF scatter, and therefore the uncertainty on the distance modulus of a
galaxy, behaves according to eqn. (3).
The change in $r/r_e$ as a function of $x$ is indicated by the solid
line. The homogeneous MBCs that would be obtained adopting fixed values
of $\Delta_\mu$ of, respectively, 0.25, 0.35 and 0.45 mag, are shown
by dashed lines. 

For individual galaxies the homogeneous MBC varies typically
between 1.05 and 1.20, respectively for galaxies on the
high and on the low end of the velocity width distribution. 
Note how the adoption of a constant value for $\Delta_\mu$
can lead to gross under-- or overestimates of the MBC. For example,
consider a fast rotator with $\log W = 2.75$, which will also be
a bright galaxy likely to be observed even at relatively large
distance. The homogeneous MBC for such an object, if measurement
errors were typical, would be $r/r_e \simeq 1.04$. On the other
hand, the adoption of a constant value for the TF scatter
of 0.35 mag (the mean value for our cluster sample), would
lead to a correction of $r/r_e \simeq 1.11$. If the galaxy were
at rest at, say $cz = 5000$ \kms, the difference in the estimated
redshift distance resulting from the two corrections would be
greater than 300 \kms : the flawed (constant $\Delta_\mu$) MBC
would produce the equivalent of a large, negative and spurious 
peculiar velocity.

For clusters, the relative error on the distance, $\Delta$, is 
roughly $N^{1/2}$ times smaller than for individual galaxies. 
As a result, except for very small samples, the homogeneous MBC
would not exceed 1.01. We will consider the impact of this small
correction in our analysis of cluster motions in Paper VIII.

\subsection{Effect of Cluster Environment}

Galaxies seen projected near cluster cores are likely to suffer
disruptive interactions with the cluster environment. It is 
therefore necessary to verify that the same UTF is satisfied
by those objects, as that representative of peripheral objects
and field galaxies. For that purpose, we perform a simple test. 
After corrections for morphological type and incompleteness
bias are applied, and after the offsets associated with cluster
motions are applied as discussed in section 6.1, the magnitude
residuals with respect to a direct TF relation of coefficients 
$(-21.01,\slop)$ are calculated. In figure 16, those residuals
are plotted versus the projected distance from cluster centers.
In panel (a), all galaxies in the {\bf in+} samples are included,
while in panels (b) and (c) galaxies have been separated between
the 12 higher density clusters (A262, A400, Cancer, Hydra, A1367,
Cen30, Coma, ESO508, A3574, A2197/9, A2634 and A2666) and the
remaining 11 lower density clusters, respectively. There is no
apparent trend in the residuals, and we judge that there
is no sufficient reason to doubt that a single UTF is satisfied
by the entire sample.

It should be noted that a dependence on cluster environment of TF 
parameters may already have been removed from the data in Figure 16, 
in the form of the morphological type correction. Because of the 
morphology--density relation, earlier spiral types are likely to be 
found in closer projection to the cluster core than late spiral types. 
It appears however that
the type correction discussed in section 3.1 applies to
early type spirals that are not found in clusters as well
(see Figure 4a), so the type correction cannot be attributed
to the effect of the cluster environment.

\section{Fitting Procedures}

For a set of $N$ data points $(x_i,y_i)$, we determine ``direct'', 
``inverse'' and ``bivariate'' forms of the linear TF relation, as well 
as a quadratic model, according to the following rules. 

The direct fit consists in the
determination of the coefficients $a_{dir}$ and $b_{dir}$ in the form

$$y(x) = a_{dir} + b_{dir} x \eqno(15)$$

\noindent
via minimization of the merit function 

$$\chi^2 = \Sigma_{i=1}^N \Bigg[{y_i - y(x_i~;a_{dir},b_{dir}) \over 
\epsilon_i} \Bigg]^2 \eqno(16)$$

\noindent
where we adopt $\epsilon_i = \sqrt{\epsilon_{y,i}^2 + \epsilon_{int}^2}$, 
the total uncertainty in $y$ arising from both measurement and intrinsic 
errors. 

The inverse fit refers to the determination of $a_{inv}$ and 
$b_{inv}$ in 

$$x(y) = -a_{inv}/b_{inv} + y/b_{inv} \eqno(17)$$

\noindent
where the merit function to be minimized is analogous to that of the
direct case, except for an exchange of $x$ with $y$, and the assignment
$\epsilon_i =\sqrt{\epsilon_{x,i}^2 + (\epsilon_{int}/b)^2}$. 

In the bivariate case, errors in both $x$ and $y$ are taken into
consideration, and we fit 

$$y(x) = a_{bi} + b_{bi} x \eqno(18)$$

\noindent
where the error used in the computation of the merit function by means of
eqn. (16) is defined as 
$\epsilon_i = [(\epsilon_{x,i} b_{bi})^2 + \epsilon_{y,i}^2 + 
\epsilon_{int}^2 ]^{1/2}$.

The three sets of coefficients $(a_{dir},b_{dir})$, $(a_{inv},b_{inv})$ and 
$(a_{bi},b_{bi})$ are computed for each subsample, 
separately for each cluster or group. For the three methods the amplitude 
of the average r.m.s. scatter figure is estimated via

$$\sigma = \sqrt {\chi^2_{min}/\Sigma_i (1/\epsilon_i^2)} \eqno(19)$$

\noindent
where $\chi^2_{min}$ is the minimum of the appropriate $\chi^2$, i.e. that 
which corresponds to the chosen solution $(a_{dir},b_{dir})$, 
$(a_{inv},b_{inv})$ or $(a_{bi},b_{bi})$. For the bivariate fit, 
we also estimate $\sigma_{abs}$, the 
mean absolute deviation in $y$. 

For the global sets resulting from the combination of all the cluster samples,
discussed in section 6, we also estimate the most likely slope of the
``primitive'' linear TF relation which the data set would follow in the absence of
scatter. We do so by producing a large number of simulated samples,
fitting each as we do for the real data sample, and keeping track of the
fit parameters. For each simulated sample, we maintain the values of the 
linewidths of the real sample, and extract random deviates of a population 
having the completeness characteristics and the average scatter behavior 
described in section 2. We loop through a set of simulated TF slopes,
and called the ``primitive'' slope that which yields the closest approximation
to the parameters fit to the real data. This value usually tracks 
closely the slope of the bivariate fit described above.

To our global sets we will also fit a quadratic relation of the form

$$y(x) = c_\circ + c_1 x + c_2 x^2 \eqno (20)$$

\noindent
where the coefficients $c_i$ are obtained by minimizing a merit function
similar to that in eqn. (16).

\section{Individual Cluster TF Relations}

Figures 17 and 18 show the TF diagrams for galaxies in {\bf in} samples
of each cluster. In Figure 17, filled circles identify galaxies in the
{\bf in} samples; the data points in this figure are not corrected
for any of the biases discussed in section 3. In Figure 18, the location
of data points is corrected for incompleteness and morphological
type bias, as discussed in sections 3.1 and 3.2, and symbols are 
replaced by error bars. The error bars refer to the ``measurement
errors'' $\epsilon_x$ and $\epsilon_y$ as defined in section 2.
The same fiducial line of slope -7.6 is plotted in each panel of
Figures 17 and 18. 

In distant clusters, we naturally tend to sample the more luminous 
galaxies. The completeness functions of all clusters,  
discussed in section 3.2, are shown in Figure 10.
Significant offsets with respect to the fiducial line
are clearly exhibited in Figure 18 by Eridanus, Ursa Major, ESO508 and 
MDL59; these are among the nearest clusters in our set. Their
samples tend to reach fainter in the LF. Because
the vertical displacements produced by motion are largest
and prevalently faintwards for these clusters, their magnitude
distributions appear to reach even deeper than they do in reality. 
The completeness functions shown in Figure 10 have been estimated
after a shift that accounts for each cluster's motion was applied.
These shifts are discussed in section 6.1.

In Table 2, for each cluster we list the three sets of fit coefficients 
$(a_{dir},b_{dir})$, $(a_{inv},b_{inv})$ and $(a_{bi},b_{bi})$ as 
defined in the preceding section. The direct and bivariate linear least 
squares fits are applied to the bias--corrected cluster samples shown in 
Figure 18. The inverse fits are applied to the cluster samples without
a correction for incompleteness bias. 
We also list estimated mean uncertainties on $(a_{bi},b_{bi})$,
and two estimates of the average scatter: $\sigma_{bi}$ (col. 11), the
scatter with respect to the bivariate relation, and $\sigma_{abs}$ (col. 12),
which is the scatter in the values of the absolute deviation about 
$(a_{bi}, b_{bi})$. The number of objects used to fit each sample, $N$,
is listed in col. (2) of Table 2. 

It should be kept in mind that an appropriate inverse fit should require
the estimate of an incompleteness bias, which will be different from the
one applied for the direct and bivariate fits, as we have discussed in
section 3.2. We have not estimated that version of the bias, and include
the inverse fits estimated without a bias correction purely for comparison
with inverse fits in other sources, which usually ignore the incompleteness
bias as well.

As mentioned before, all galaxies comprising the
{\bf in} samples were assigned the same distance, that obtained from the
cluster systemic redshift. This assignment was made on the basis of angular
and kinematical information which is not completely unambiguous. It is
possible that some of the objects assigned cluster membership are decoupled
from the virialized cores and partake of pure Hubble flow. If a significant 
fraction of galaxies in our samples suffered from that misclassification,
an important bias could be introduced in both the individual cluster TF
relation and in the global template to be constructed by merging  all
cluster samples. This bias would result because, especially for more 
distant clusters, interlopers in the foreground would more often be 
included than those in the background. In Figure 19 we display the 
magnitude residuals from a TF relation of slope -7.6, after 
forcing the mean residual to be zero for each cluster sample. This is 
done for each of the {\bf in} cluster samples. Residuals are plotted 
versus the logarithm of the 
redshift of each galaxy, to which the logarithm of the systemic redshift 
of the cluster is subtracted, so that in each plot the center of the cluster
corresponds to the point (0,0). Galaxies which are freely expanding with 
Hubble flow but were erroneously assigned cluster membership would be found
to scatter about a line of slope 5, as indicated by the dashed line, in
Figure 19. Any preferential alignment of data points with that line
would indicate contamination of the cluster sample with interlopers.
While it may be possible to question a few data points, by and 
large the cluster diagrams yield 
little evidence of any systematic contamination. Fig. 20 displays all {\bf in} 
samples merged together in a similar representation, illustrating that 
no significant forward--backward interloper bias affects globally our 
{\bf in} samples.

\section{TF Template Relation}

The individual, bias--corrected TF data sets of each cluster can
be combined to obtain a TF template relation. We will refer to the
result of the combination as the global set. We first describe the
procedure adopted in carrying out the cluster combination. We then
derive the parameters that best describe the template relation fit
to the global set and discuss the extent to which it approximates 
a desired UTF.

\subsection{Cluster Combination}

For inclusion in the global sample, the absolute magnitude of the 
$j$--th galaxy in the $k$--th cluster is corrected in the mode

$$y^c_{j,k} = y_{j,k} + \beta_{typ}({\rm type}_{j,k}) 
+ \beta_{inc,k}(x_{j,k}) - \Delta y_k \eqno(21)$$

\noindent
where $y_{j,k}$ is the magnitude listed in Table 2 of Paper VI, 
$\beta_{typ}$ is the morphological type correction described in section 
3.1, $\beta_{inc,k}$ is the incompleteness bias shown in Figure 10, which 
is a different function for each cluster, and $\Delta y_k$ is an offset, 
also specific to each cluster sample. The only ingredient we have not 
discussed so far is the offset $\Delta y_k$. This set of parameters is 
introduced to represent the motion of each cluster, with respect
to the reference frame defined by the template relation. 

For a linear TF of slope $b$, the TF zero point of the cluster $k$ will 
be $a_k = a_\circ + \Delta y_k$, where $a_\circ$ is a constant. We
simultaneously determine $b$, $a_\circ$ and the set of offsets 
$\Delta y_k$ by minimizing the quantity

$$\chi^2 = \Sigma_{j,k} \Bigg[{y^c_{j,k}  - a_\circ - b x_{j,k}  \over 
\epsilon_{j,k}} \Bigg]^2 \eqno(22)$$

\noindent
with the condition that

$$\Sigma_{k} N_k \Delta y_k /\Sigma_{k} N_k = 0 \eqno (23)$$

\noindent
where the sum is intended over a subset of ``distant''clusters,
the choice of which is discussed in the next section, and $N_k$ 
is the number of galaxies for the $k$--th cluster sample. The
errors $\epsilon_{j,k}$ are as defined after eqn. (18). In the
case of a quadratic fit (eqn. 20), a similar minimization
process is followed, with the difference that an additional
coefficient is simulataneously determined.

Eqn. (23) defines our ``inertial frame'', i.e. it implies that
the set of ``distant'' clusters involved in the summation
defines a mean null velocity with respect to the CMB. We return to
this point in section 6.2.

In Table 3, we list values of $\Delta y$ for each cluster sample: 
in cols. (3--5) after linear TF fits and in cols. (6,7) after quadratic
fits. In col. (3), the offsets $\Delta y$ are
computed after applying a correction for morphological type,
but not one for incompleteness bias, to the data. In cols. (4) and (5) the 
incompleteness bias correction is diversely applied: in col. (4),
it is computed assuming an LF with a power law
exponent $\alpha = -0.50$, while the value adopted for the computation
of offsets in col. (5) is $\alpha = -0.85$ (see section 3.1.1 and Figure 8).
In cols. (6) and (7), offsets were computed after fitting a quadratic
TF relation; incompleteness bias corrections for columns (6) and (7)
are the same as adopted for cols. (4) and (5), respectively.
An estimate of the uncertainty of the last two figures of each value 
is given between brackets, for the values tabulated in col. (4);
similar uncertainties apply to each column. The uncertainties were 
estimated by taking the largest number between the mean r.m.s. scatter 
of the fit to the cluster sample, $\sigma_{bi}$ (tabulated in col. 11 
of Table 2) and a value of 0.35 mag, and dividing 
that value by the square root of the number of objects in the given
cluster sample. Uncertainties in the cluster systemic redshift
were not included in this estimate. The values listed in Table 3
will be used in paper VIII in the analysis of the cluster peculiar 
velocity field.

The global sample obtained by combining individual cluster samples
as described by eqn. (21) yields a TF relation of slope and 
zero point that are corrected for incompleteness and morphological type 
bias and for the motions of individual clusters, to the extent that the 
``distant'' cluster subset can be assumed to yield a null mean velocity 
with respect to the CMB. Because the global sample is the sum of 
many incomplete cluster samples, it does not populate the TF plane in 
the same manner that a complete, volume limited sample would. While its
slope and zero point should be expected to mimick those of an unbiased
sample with the same sky and velocity distribution characteristics, its 
measured scatter will underestimate that of an unbiased
sample, for the reasons discussed in section 3.2.2 and illustrated in 
Figure 6(c). We will discuss that effect in a more quantitative way in
section 6.3.

\subsection {The Cluster Reference Frame}

The bias corrections that were applied to each of the cluster samples 
help produce a global template with a slope
which is as close as we are likely to get to that of the ``primitive''
UTF. Ideally we would wish the zero point to identify with that of a
reference frame at rest with respect to the CMB. Since the zero
point is specified in magnitudes, and since it is fair to assume that
the cluster peculiar velocity distribution function does not
depend on geocentric distance, the TF zero point is best defined
if the cluster set used to specify it is a distant one: for 
$V_{pec}/cz \ll 1$, the corresponding magnitude offset is
$\Delta m \simeq -2.17 V_{pec}/cz$.

Let the cluster velocity distribution function be characterized 
by a r.m.s. one--dimensional peculiar velocity $<V^2_{pec}>^{1/2}$.
Then the average of the velocities of $n$ randomly chosen clusters, 
placed at a mean redshift $<cz>$, would 
yield a most probable magnitude offset from the UTF of
 
$$|\Delta m|_{exp} \sim 2.17 <V_{pec}^2>^{1/2} <cz>^{-1} N^{-1/2}  
\eqno(24)$$

\noindent
Bahcall and Oh (1996) report a value of  $<V_{pec}^2>^{1/2}=293\pm28$ \kms.
For $N=16$ and $<cz> = 6000$ \kms, $|\Delta m|_{exp} \sim 0.03$ mag.
Since clusters in our sample are not randomly distributed, e.g.
N383, N507 and A262 are in the Perseus--Pisces supercluster, A1367
and Coma are both in the Coma supercluster, A2197 and A2199 are
similarly associated, etc., the actual number of clusters with  
uncorrelated velocities is less than 24. In addition,
it is wise to remove those clusters that are at $cz<3000$ \kms or
that exhibit large motions, from the set that will be used to determine 
the TF zero point, in order to reduce the kurtosis of
the distribution of magnitude offsets. In Table 1, we have labelled
with an asterisk the 14 clusters that have been chosen 
to define the TF zero point. Their mean redshift is about 6,000 \kms
and we shall assume conservatively that they represent about 9 independent 
large scale aggregates.

A set of cluster galaxies will allow the definition of a TF template
with a statistical accuracy that improves with the number of objects
involved in the TF relation. For example, for a mean scatter of about
0.35 mag, the sample of about 250 {\bf in} 
galaxies in the combined 14 reference clusters will yield a TF 
zero point with a statistical error of 0.03 mag. This is independent
on whether the cluster set represents a system globally at rest
with respect to the CMB or not. Using eqn. (23), we can estimate
that the reference cluster sample approximates
the TF zero point, identifying rest with respect to the CMB, to within
approximately $\sim 0.035$ mag. Combining these two figures in quadrature,  
we obtain an expectation of kinematic zero point accuracy for the global
{\bf in} sample of 0.045 mag. For the {\bf in+} sample, the expectation
improves slightly, to 0.040 mag. The sparse number and uneven sky coverage
of the distant cluster samples advises caution, however, suggesting that
somewhat larger uncertainties may apply. We expand on this 
analysis in Paper VIII.

\subsection {Fits to  Cluster Templates}

Fig. 21(a) displays the global template obtained by combining the
{\bf in} cluster samples. In Figure 21(b) the analogous combination
is displayed for the {\bf in+} samples and in Figure 21(c) we show only the
measurement error bars $(\epsilon_x, \epsilon_y)$ for the objects
in Figure 21(b).

In Table 4 we give the results of linear fits to the global sets.
First, we display the results for the global {\bf in} and {\bf in+}
sets, {\it without correcting cluster samples for the incompleteness
bias}; the fit parameters are given by the first two rows of the Table.
Then we display fits to several subsets of bias--corrected samples,
separately for the two cases in which the bias computation was 
carried out using a faint--end power law exponent of the luminosity
function $\alpha = -0.50$ and $\alpha = -0.85$. For each value of
$\alpha$ we fit several subsets of the data, as described in col. (1)
of Table 4, thresholding the data by inclination, type, velocity
width and $2.5\sigma$--clipping. In each case, we also estimate the 
most likely slope of the ``primitive'' TF relation which would yield the 
given set of fits, $b_p$, listed in col. (13). As mentioned 
in section 4, the primitive relation is usually well approximated by 
the bivariate fit. 
The formal errors on the slope of the fits are typically on the
order of 0.15, or 2\%, while those on the zero point are on the
order of 0.02 mag. For the bias--corrected samples, no inverse fits
are computed.

Among the fits to data subsets in Table 4, the ones to objects of
type Sb and earlier yield significantly lower slopes than the other
subsets. This may be an indication of the fact that our type correction
discussed in section 3.1 is inadequate. As we discussed in that section,
however, the completeness of early type objects may differ from that
of those of later type, and the difference in slope may partly reflect
that aspect. As the number of early type objects included in TF samples
increases, it will be possible to assess the character of the applicable
internal extinction laws, TF relations, etc. For the moment, we choose
to apply the simplest of corrections, as indicated in section 3.1.

As expected, the correction for the incompleteness bias dims the 
zero point and steepens the TF relation. The effect is stronger
as the value of the LF exponent $|\alpha |$ increases, 
because that implies that samples become progressively more incomplete, 
and thus require a larger bias correction.

The {\bf in} samples exhibit a slightly steeper
slope than the {\bf in+} sample, but the difference is of marginal
significance at best.

It is also useful to point out that the cluster offsets $\Delta y$
are not very sensitive to the choice of the adopted LF,
or even to whether the incompleteness bias correction is applied at all.
That is because the cluster offsets $\Delta y$ are affected by the
{\it relative differences} of the incompleteness bias corrections,
averaged over all galaxies in a cluster sample, rather than by their 
actual amplitudes. 

Finally, in Table 5 we give fit coefficients for the quadratic
relation (20), for a more restricted set of samples. The statistical
significance of the difference between the linear and quadratic
fits is marginal for the cluster set. However, it appears that 
further inclusion of dwarf and irregular galaxies, extracted from
the low luminosity end of the LF, would drive the TF slope steeper
and introduce noticeable nonlinearity (Hoffman and Salpeter 1996).
While the use of such objects on peculiar velocity studies is of
very limited usefulness, due to the large associated scatter, a
sample which includes them would be poorly served by a strictly linear
TF relation.

\subsection {Underestimate of Scatter in the Template}

In section 3.1 we discussed how the incompleteness bias affects
not only the slope and zero point of the TF relation, but also its
measured scatter. The effect is one of reducing the apparent scatter
near the faint end of the TF relation, as shown in Figure 6(c).
Before combining cluster sets, each galaxy is shifted by the incompleteness
bias correction, which produces a TF relation in which bias for the
slope and zero points have been formally removed. The scatter, however,
remains underestimated. The range of velocity widths over which the
underestimate occurs varies from cluster to cluster, shifting to lower
widths for the more nearby clusters. When the clusters are combined to
obtain the global set, mild underestimates of the scatter result throughout
the lower half of the velocity width range. We have obtained a Monte Carlo
simulation of the underestimate, by producing the combination of the
cluster samples, simulated according to the incompleteness characteristics
of each as described in section 3, and repeating the simulation a large
number of times (800 in our case). Figure 22 shows the results of this 
simulation, and illustrates
that the underestimate is quite mild. If we fit the data in Figure 21(b) and 
measure the scatter as a function of velocity widths, that measure is always 
at least 90\% as large as the real scatter that would result if we had dealt
with complete samples. Thus, the average, measured scatter values
tabulated in Table 3 underestimate the true scatter by a small amount. The 
bias corrected values of the TF scatter, for each tabulated sample, exceed
those measured by about 0.02 mag, and are not tabulated separately.
The bias correction was however taken into account in section 2, in the 
analysis of the properties of the scatter in the TF plane (see fig. 1).
The smooth line fitted to the simulation data in Figure 22, which parametrizes
the underestimate of the scatter as a function of velocity width, was
used to correct the location of the unfilled symbols in Figure 1 to that
of the filled circles and crossed squares, which represent estimates of
the bias--corrected total scatter of the TF relation.

\section{TF Residuals}

A simple test of adequacy of a TF template relation consists in the 
investigation of the behavior of residuals. Systematic trends in the amplitude 
and scatter in the residuals have already been used in section 2 (Fig. 4),
to obtain an understanding of the components in the TF relation error
budget; in section 3.1 (Figs. 5 and 6) to infer morphological type 
differences in TF offset and slope; in section 5 (Figs. 19 and 20) to
infer that cluster samples are not seriously contaminated by fore--
and background objects; and in section 3.3 (Fig. 24) to conclude
that no catalog edge bias effects are present in our data. The only
significant evidence for a trend in the residual distribution was noted 
in the their dependence on morphological type, which was subsequently
modelled and resulted in a morhological type correction to the data.

In Fig. 23, we plot the magnitude residuals of individual galaxies in
the {\bf in+} global sample, versus angular size (panel $a$) and 
apparent magnitude (panel $b$). The residuals are computed with respect
to the bivariate fit of slope \slop. We plot individual galaxies as 
small unfilled triangles and running averages as large, solid circles.
We note that for intermediate values of the angular size, or of the
magnitude, residuals are flat and there is no trace of a trend.
However, at the two ends of the distributions residuals deviate
slightly from zero, on the average. At first, this may be thought of as 
the signature of a problem in our treatment, while in reality the
observed weak deviation of the average deviations is exactly what
should be expected. Consider the simulated cluster sample shown in
figure 5. The few faintest galaxies in the sample tend to be below the
TF line, and the few brightest tend to be above the TF line; the amount
by which this deviation occurs depends on the amplitude of the TF
scatter. The effect is greatly reduced (but not cancelled) if residuals
are estimated with respect to the inverse fit. For every cluster
sample, the faint end occurs at different absolute magnitudes, but
generally at the same apparent magnitude, so that the cluster combination
does not cancel the effect. Simulations of the effect show that the
expected amplitude of the main residuals should be about as large as
observed.

In Figure 24(a), we show the residuals with respect to the above mentioned 
bivariate 
relation, plotted versus the redshift of each individual galaxy (rather 
than that of the cluster as a whole, thus reducing the degree of crowding
of the data points in the plot). Reassuringly, the average residual
remains null through the whole redshift range. In Figure 24(b), residuals
are plotted versus disk inclination angle to the line of sight, verifying
that the adopted correction schemes which depend on inclination are
adequate.

It appears that no significant, unexpected biases, related with observed 
parameters of sample objects, affect our global cluster set.

\section{Summary}

We have presented I band photometry and spectroscopic data of 782 spiral 
galaxies in 24 clusters of galaxies, garnered from our own observations 
and from material in the public domain. Of those, 555 galaxies are cluster 
members or lie close enough to the clusters' peripheries, to be useful in a 
study geared at obtaining a TF template relation. We have divided the galaxies 
in two groups: an {\bf in} group of 360 galaxies, thought to be bona fide 
cluster members, and an {\bf in+} which includes in addition objects that are 
peripheral to the clusters.

We analyzed in detail the characteristics of the {\it scatter in the TF 
relation}. The average total scatter is about 0.35 mag; its amplitude however 
varies between about 0.25 mag among fast rotators and more than 0.40 mag among 
the slowest rotators in our sample. Measurement errors contribute to this trend,
in the sense that they tend to be globally larger for less luminous, slow 
rotators. However, we found that uncertainties associated with measurements and 
with corrections applied to the observed parameters cannot fully account for the total observed scatter. An additional component, which we refer to as ``cosmic'' or ``intrinsic'', is necessary. This component may result from a variety of 
sources, such as asymmetries in the spiral disks' light and velocity fields and 
differences in the formation histories of galaxies. The intrinsic scatter 
component varies with galaxy luminosity, or velocity width, being larger --- 
perhaps as large as 0.35 mag --- for the slowest rotators. The intrinsic scatter among the most luminous, fastest rotators may be 0.20 mag or even smaller. We 
found that the systematically changing character of the TF scatter has 
important effects in the estimate of bias corrections, and that the assumption 
of constant values for the scatter can easily lead to spurious results. 
In addition, the variety of sources of scatter weakens the merits of adopting 
the inverse form of the TF relation as a bias--free device. 

We have discussed in detail the impact of {\it galaxy morphology}, 
{\it environment and bias on the TF relation}. 

We found that, even when I band photometry is utilized, galaxies exhibit 
differences in their TF properties that correlate with {\it morphological type}. 
A correction is necessary to galaxies of earlier spiral types, in order to 
make their use compatible with that of the Sbc and Sc galaxies that form the 
bulk of our sample. We have used the simplest of corrections, that of a simple 
additive term, in order to bring the earlier types to consistency with the 
later ones. This is a conservative choice that may turn out to be inadequate. 
With varying type, it is quite possible that variations in the light--to--mass 
ratios and internal extinction properties may lead to changes in the TF slope. 
Since the number of objects of type earlier than Sb in our sample is 
relatively small, and the behavior of Sb galaxies appears only marginally 
different from that of later types, we have chosen not to apply more elaborate 
correction schemes until the size of TF samples of earlier types increases, and 
independent determinations of mean properties become more reliable.

The effect of {\it incompleteness} was analyzed next, and a Monte Carlo technique was
described for the computation of the associated bias. We discussed the sensitivity
of the bias estimates on the scatter of the TF relation, the shape of the luminosity
function of the galaxy population and the slope of the TF relation. An accurate
determination of the scatter properties is necessary for a reliable estimate of
the bias, as the scatter amplitude is its main driver. 

We then analyzed other possible sources of bias, namely: (a) that arising from the
nearness of sample galaxies to the limit of the catalog from which the objects
are extracted, which in our case is negligible; (b) the effect of cluster depth
along the line of sight, also a very minor effect; (c) the dependence of TF
properties on galaxy location in the cluster, which we found to be undetectable
in our sample. Lastly, we considered the homogeneous Malmquist bias. This effect 
is relatively unimportant for cluster samples but it can make substantial corrections 
necessary for individual galaxy distances. In general, it is modulated by the
density function of the galaxy distribution and it is most often applied as an 
`inhomogeneous' Malmquist bias correction. In either case, we found that the 
Malmquist bias amplitude is strongly affected by the variable character of the 
TF scatter, and that the application of the correction in its `standard' form
can lead to grossly deviant estimates of a galaxy's TF distance.

Finally, we obtained a {\it global template TF relation} by combining the 
bias--corrected TF data of the 24 clusters. In the process, we estimated 
relative offsets of each cluster from the reference frame formed by the set of 
the 14 most distant clusters, which defines the best estimate of null velocity
with respect to the cosmic microwave background radiation field that can be 
garnered from our sample.

The TF relations thus obtained will be utilized elsewhere in the analysis of the
cluster motions and of the peculiar velocity field of individual field galaxies.

\acknowledgements
It is a pleasure to thank Dr. M. S. Roberts for carefully reading this 
paper and making a number of suggestions that improved it substantially.
The results presented in this paper are based on observations carried out at
the Arecibo Observatory, which is part of the National Astronomy and 
Ionosphere Center (NAIC), at Green Bank, which is part of the National Radio 
Astronomy Observatory (NRAO), at the Kitt Peak National Observatory (KPNO), the 
Cerro Tololo Interamerican Observatory (CTIO), the Palomar Observatory (PO), 
the Observatory of Paris at Nan\c cay and the Michigan--Dartmouth--MIT 
Observatory (MDM). NAIC is operated by Cornell University, NRAO  by  
Associated Universities, Inc., KPNO and CTIO by Associated Universities 
for Research in Astronomy, all under cooperative agreements with the National 
Science Foundation. The MDM Observatory is jointly operated by the University 
of Michigan, Dartmouth College and the Massachusetts Institute of Technology 
on Kitt Peak mountain, Arizona. The Hale telescope at the PO is operated by 
the California Institute of Technology under a cooperative agreement with 
Cornell University and the Jet Propulsion Laboratory.  
This research was supported by NSF grants AST94--20505 to RG, AST90-14850 and 
AST90-23450 to MH and AST93--47714 to GW.


\begin{figure} 
\caption{A graphic summary of the TF error budget. The two dotted lines
represent the uncertainties arising from measurement errors and from the
correction applied to observed parameters: $\epsilon_y$ is the error on
the magnitude and $\epsilon_x$ is the error on the velocity widths; the 
latter is multiplied by the TF slope so that it can be expressed in mags.
$\epsilon_x$ and $\epsilon_y$ are computed using not only the galaxies
in the cluster sample (SCI), but also those in a larger field sample (SFI),
providing a more accurate estimate. The solid line labelled $\epsilon_m$
is the total measurement error, as given by eqn. (10), and represents
the total contribution of measurement and correction errors to the TF
scatter. It is approximated by $[(b\epsilon_x)^2 + \epsilon_y^2]^{1/2}$,
but not exactly matched by it, because the errors $\epsilon_x$ and $\epsilon_y$
are partially correlated.
The three dashed lines are the results of sums in quadrature of $\epsilon_m$ 
and an ``intrinsic'' scatter figure of, respectively, 0.20, 0.25 and 0.30 
mag, bottom to top. The unfilled symbols are the measured, mean absolute
deviation (squares) and standard deviation (circles) of the residuals
from a linear TF relation. These values underestimate the scatter due
to the incompleteness bias discussed in section 3.2. Finally, the solid
circles and the crossed squares are respectively the incompleteness 
bias--corrected values of the standard deviation and mean absolute 
deviation: these represent the true total scatter of the TF relation.
The heavy solid line is the combination in quadrature of the measurement
error, $\epsilon_m$, and the intrinsic scatter $\epsilon_{int}$ given
by eqn. (11).}
\end{figure}

\begin{figure} 
\caption{Relative contributions of various components to the total
measurement error $\epsilon_m$, for two representative cases.  Refer
to eqn. (10) and section 2.2 for details. Panel (a) applies 
to objects with fairly well determined width $\epsilon_W/W = 0.02$ and a large 
value of $\gamma =1$, as it might correspond to a fairly bright and rapidly 
rotating galaxy. Panel (b) applies to objects with $\epsilon_W/W = 0.07$ and 
a relatively small value of $\gamma = 0.6$, as it might correspond to a 
relatively faint galaxy,
with a velocity width determination of less than average quality. In 
both cases, the error in the determination of the magnitude (before applying
extinction corrections) is assumed to be 0.05 mag.}
\end{figure}

\begin{figure} 
\caption{TF diagrams of galaxies in the {\bf in+} cluster samples; each
cluster sample has been shifted by an amount $\Delta y$ as computed
in section 6, correcting for the cluster peculiar motion with respect to 
the overall sample. Different symbols are used for different types, as 
labelled. Panel (a) displays data uncorrected for any bias, while in panel 
(b) data points have been shifted by the amount indicated by eqn. (8) and
Table 1. In each panel, the solid line identifies the fit to the late
spirals.}
\end{figure}

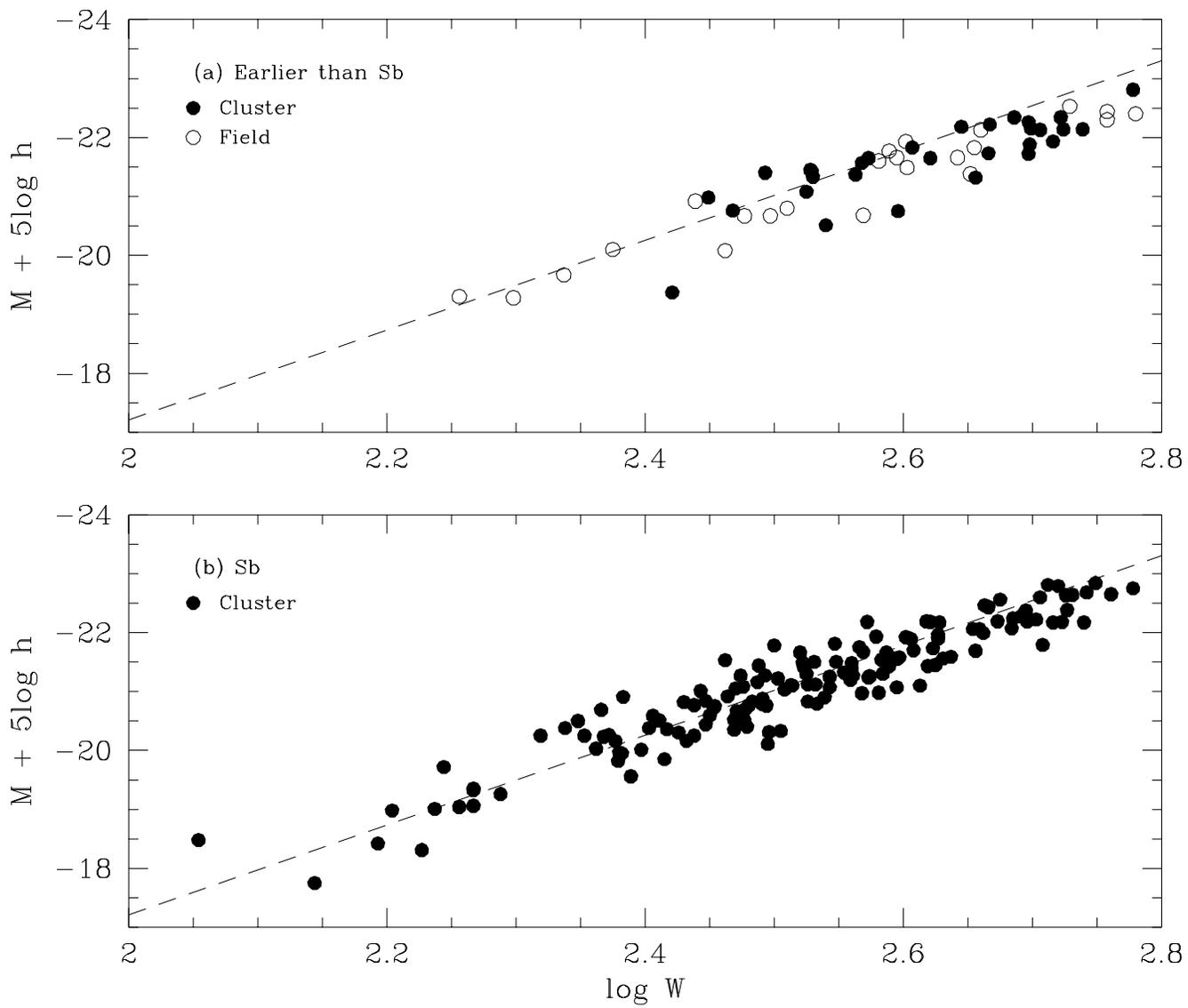
\begin{figure} 
\caption{TF diagrams of galaxies of type earlier than Sb (panel a) and Sb (panel b).
Data points are uncorrected for a morphological type or any other kind
of bias. The dashed line in both panels is the fiducial TF relation
fit to late spirals (Sbc and Sc), as shown in Figure 3; the solid lines
identify the adopted TF relation for the two type groupings displayed
here, showing the departure from the fit to later spirals. In panel
(a), early type galaxies from our field galaxy sample (SFI) have
been added (unfilled symbols), in order to increase the statistical 
significance of the adopted correction.}
\end{figure}

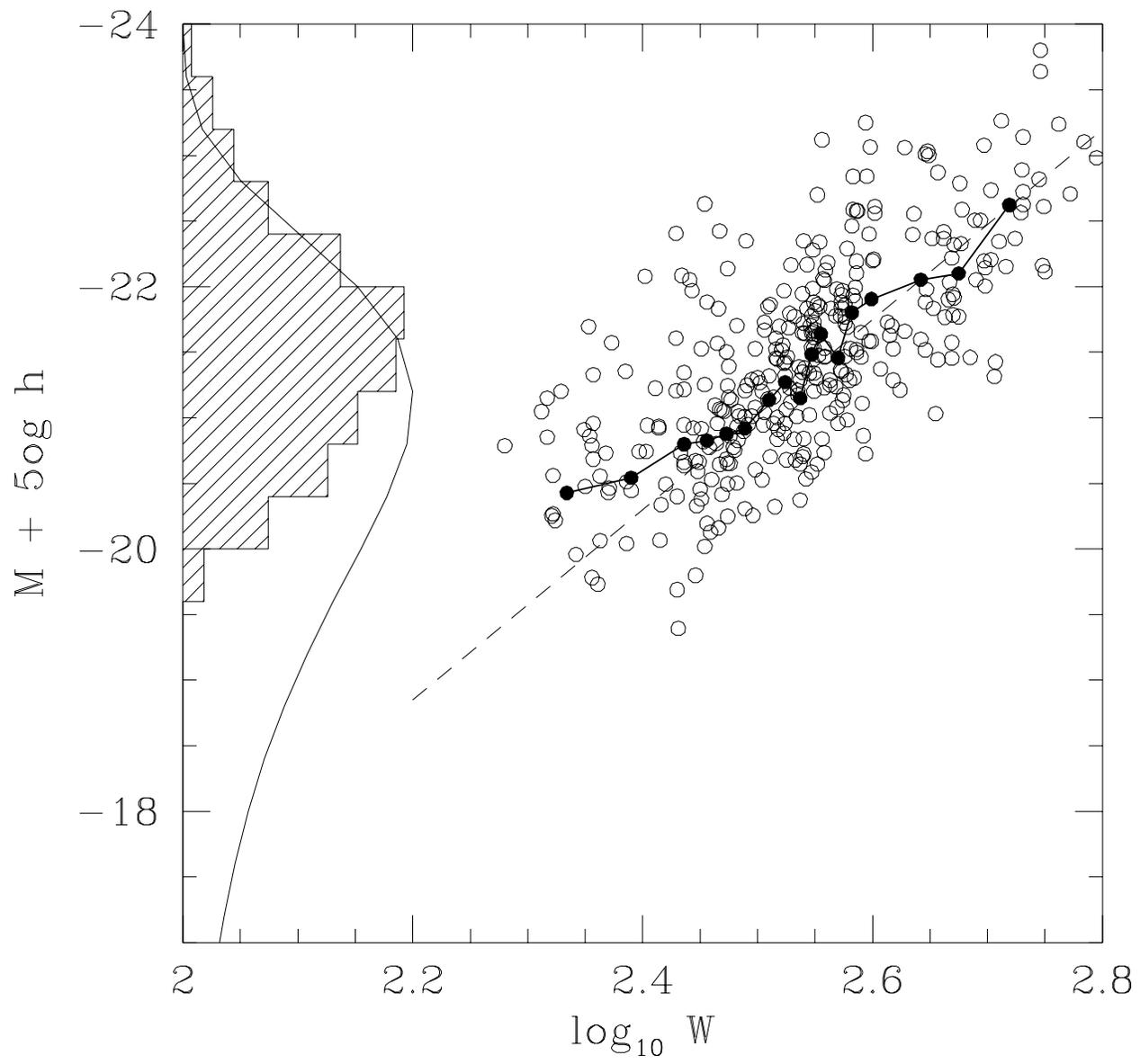
\begin{figure} 
\caption{Simulation of cluster incompleteness bias. A cluster sample
following a UTF as indicated by the dashed straight line is generated;
the individual data points are represented by unfilled circles. They
are random deviates of the LF shown by the solid curve
plotted along the magnitude axis, and extracted to have completeness
characteristics as shown by the shaded magnitude histogram. The
completeness function of the cluster is the ratio of the histogram
to the LF. A running average of the magnitudes of objects in the sample,
contained within non--overlapping bins of velocity width, is indicated 
by the solid symbols connected
by the heavy solid line: as the sample becomes progressively more
incomplete towards smaller velocity widths, the mean magnitude
increasingly departs from the UTF. Note that the scatter of the
data has been exaggerated to an amplitude twice as large as that 
indicated by eqn. (3).}
\end{figure}

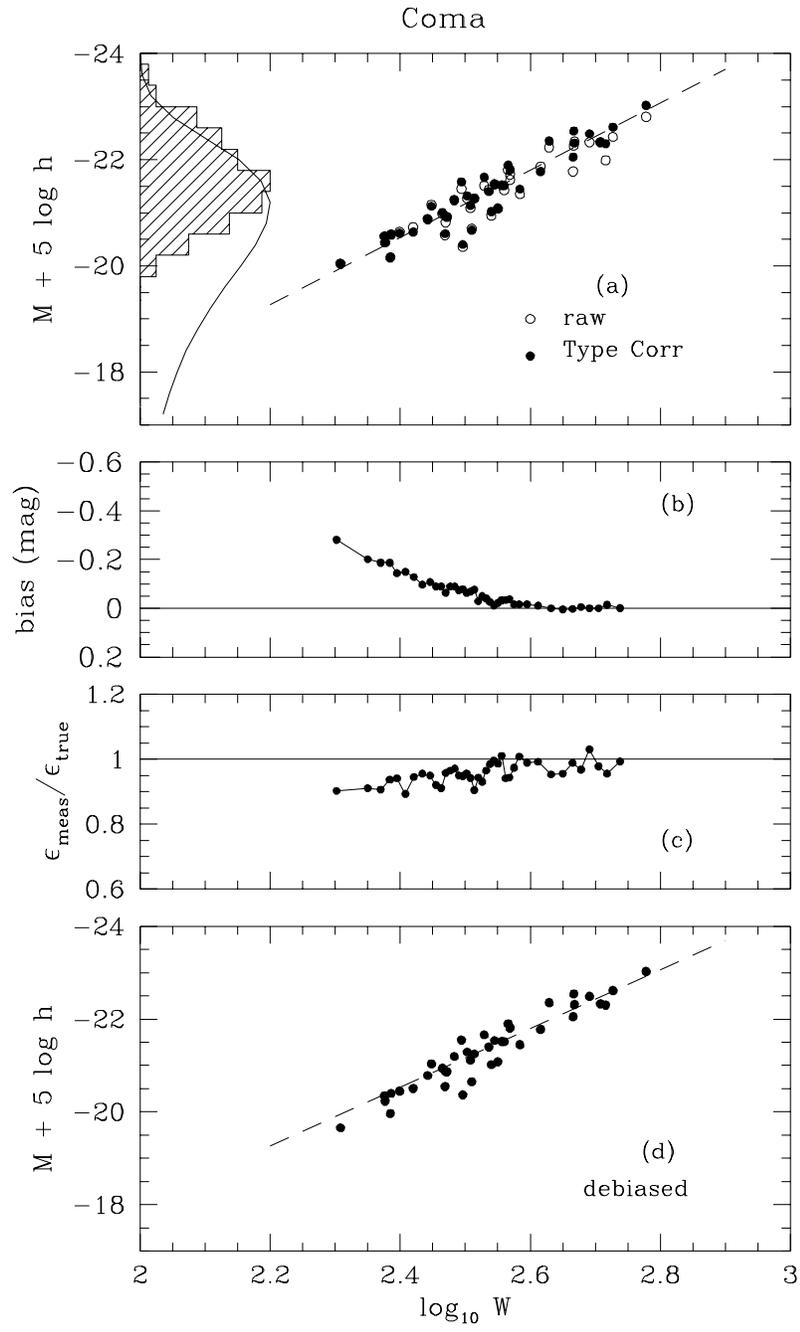
\begin{figure} 
\caption{Bias correction for the Coma cluster.
 (a) unfilled symbols
represent the raw data for galaxies in the {\bf in+} sample of the
cluster. After the application of the morphological type correction
discussed in section 3.2, the points shift to the locations of the
filled circles. The dashed straight line identifies a direct
fit to the morphological type--corrected data. The luminosity
function of the galaxy population from which the sample is assumed to
be extracted is shown as a solid line plotted along the vertical axis,
together with the shaded histogram of magnitudes for the sample. The
completeness function of the sample is the ratio between the histogram
and the LF. 
(b) Amplitude of the bias associated with the incompleteness of the
sample, as described in section 3.2. This was computed by producing
$n_{times}=800$ independent simulations of the Coma sample.
(c) Effect of the incompleteness bias on the TF scatter: the ratio
$\epsilon_{meas}/\epsilon_{true}$ is between the mean measured scatter
in $n_{times}=800$ independent simulations of the Coma sample and the
actual value of the scatter input with the model. The measured scatter
between $\log W = 2.3$ and 2.5 is underestimated by about 10\% .
(d) Coma sample data after the bias correction shown in panel
(b) has been applied, point by point. The dashed line is the same
as that in panel (a); it has been added to aid eye estimate of the
amplitude of the incompleteness bias correction.}
\end{figure}

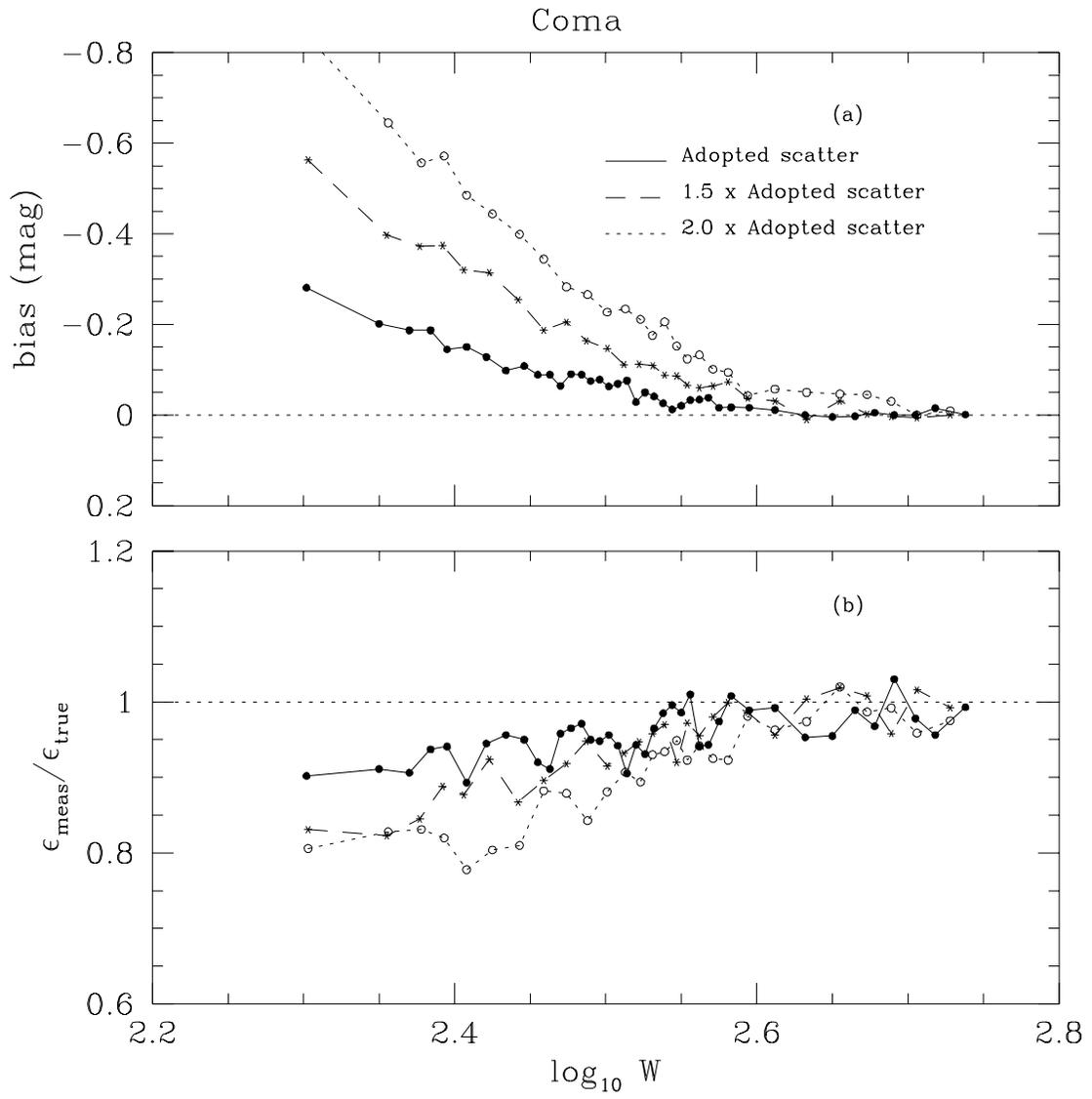
\begin{figure} 
\caption{Dependence of the incompleteness bias correction for the Coma 
{\bf in+} sample on the
amplitude of the scatter. Analogous plots to those in Figure 6b and 6c
have been produced, assuming different amplitudes of the TF scatter.
The solid lines were computed for a TF scatter amplitude as estimated
in section 2 and parametrized by eqn. (7); they are the same as shown 
in Figure 6. The long--dash and the dotted lines assume scatter amplitudes 
respectively 1.5 and 2 times larger than that given by eqn. (3).}
\end{figure}

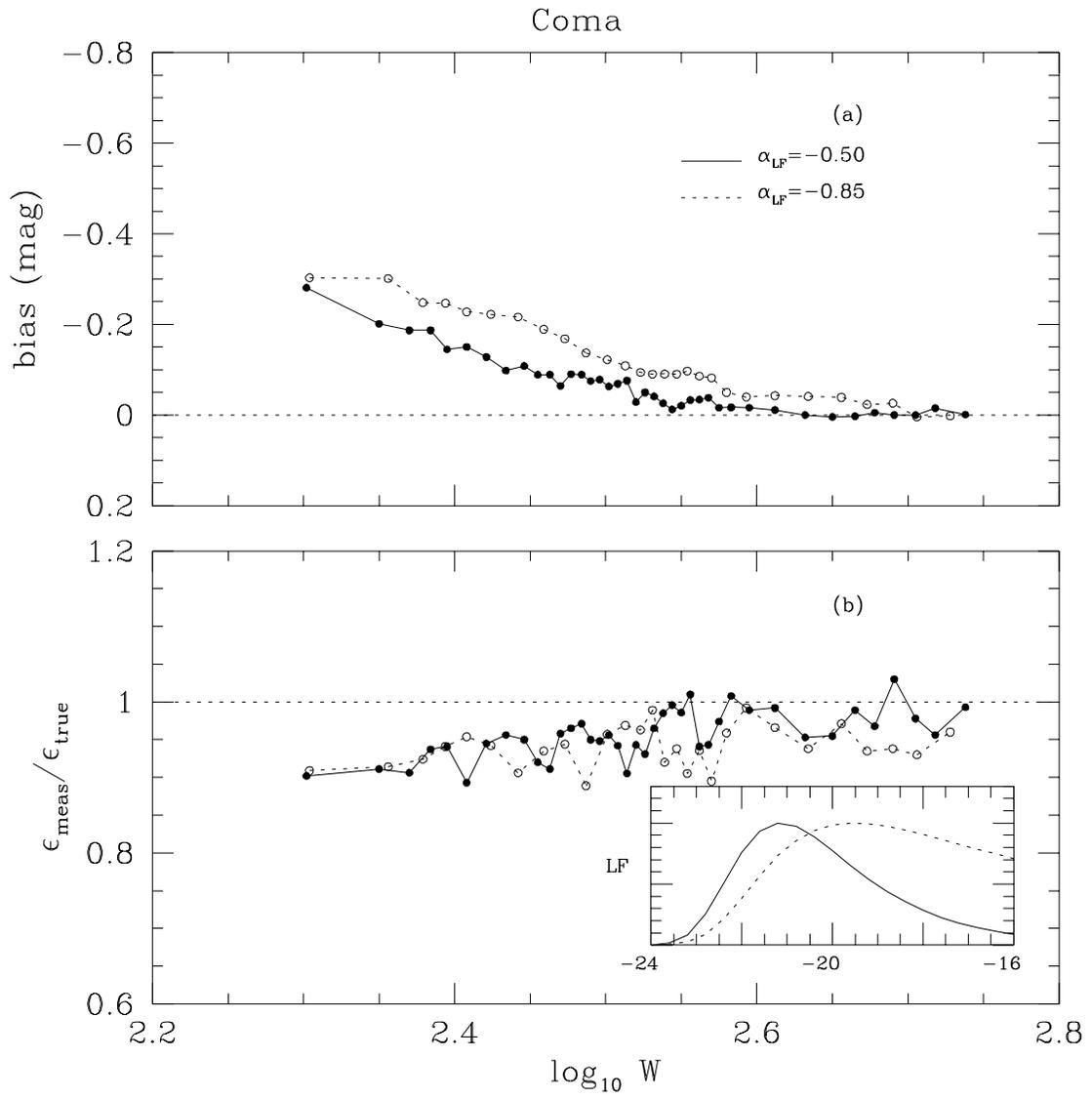
\begin{figure} 
\caption{Dependence of the incompleteness bias correction for the Coma 
{\bf in+} sample on the
shape of the LF. Analogous plots to those in Figure 6b,c 
and Figure 7 have been produced, assuming different shapes for the
LF of the population from which the sample is supposed 
to be extracted: the two funtions are shown as an inset in panel (b);
they differ by the value of the exponent $\alpha$ of the power law part
of the Schechter representation of the LF, the solid
line refers to $\alpha = -0.50$ and the dotted one to $\alpha -0.85$. 
Correspondingly, the incompleteness bias is shown for the two cases as
solid (filled symbols) and dotted (unfilled symbols) lines.}
\end{figure}

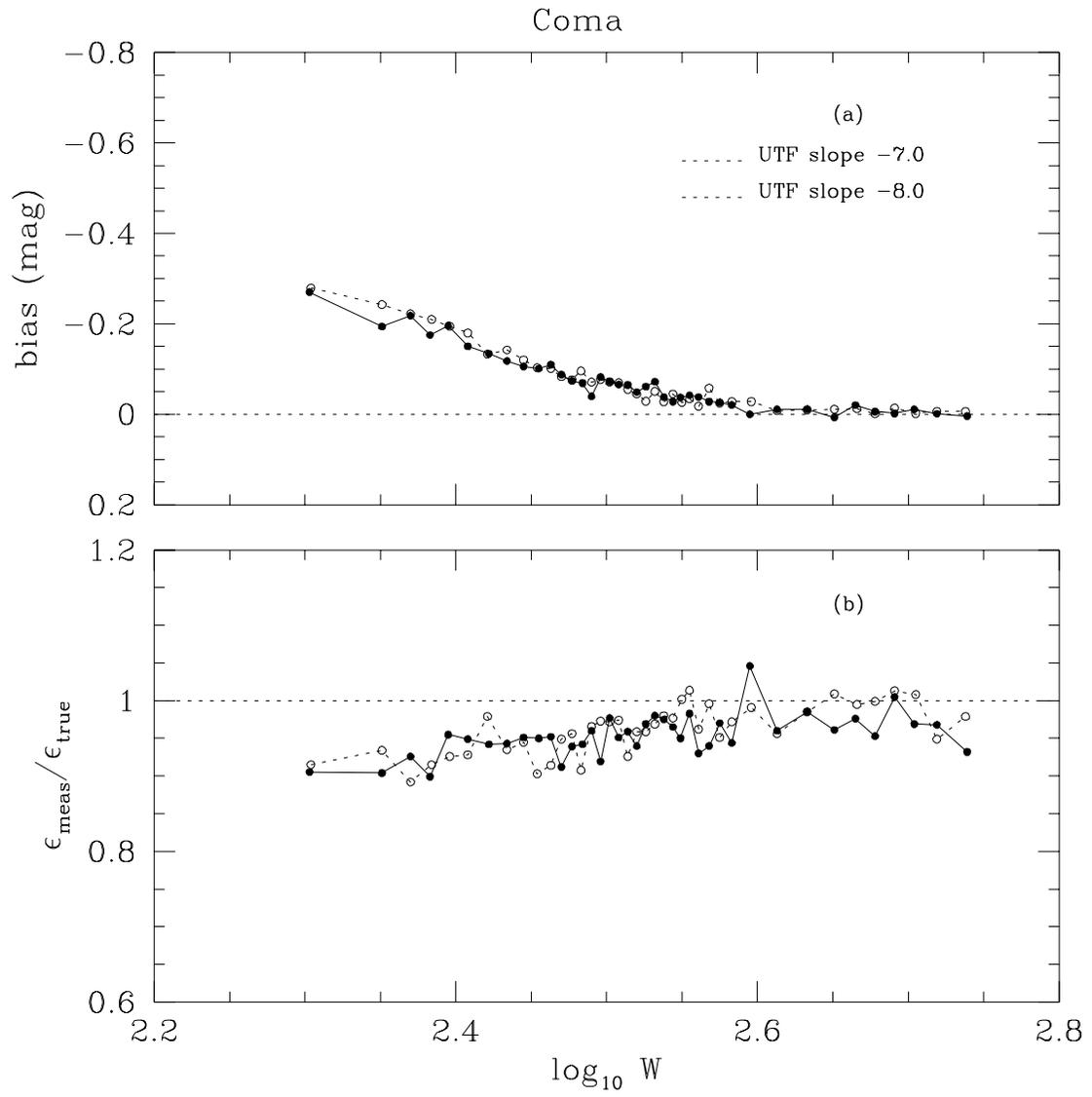
\begin{figure} 
\caption{Dependence of the incompleteness bias correction for the Coma 
{\bf in+} sample on the slope of the TF fit. Analogous plots to those in 
Figure 6b,c, Figure 7 and Figure 8 have been produced, assuming slopes of
respectively $-7.0$ (solid line, filled symbols) and $-8.0$ (dotted 
line, unfilled symbols) for the TF relation.}
\end{figure}

\newpage

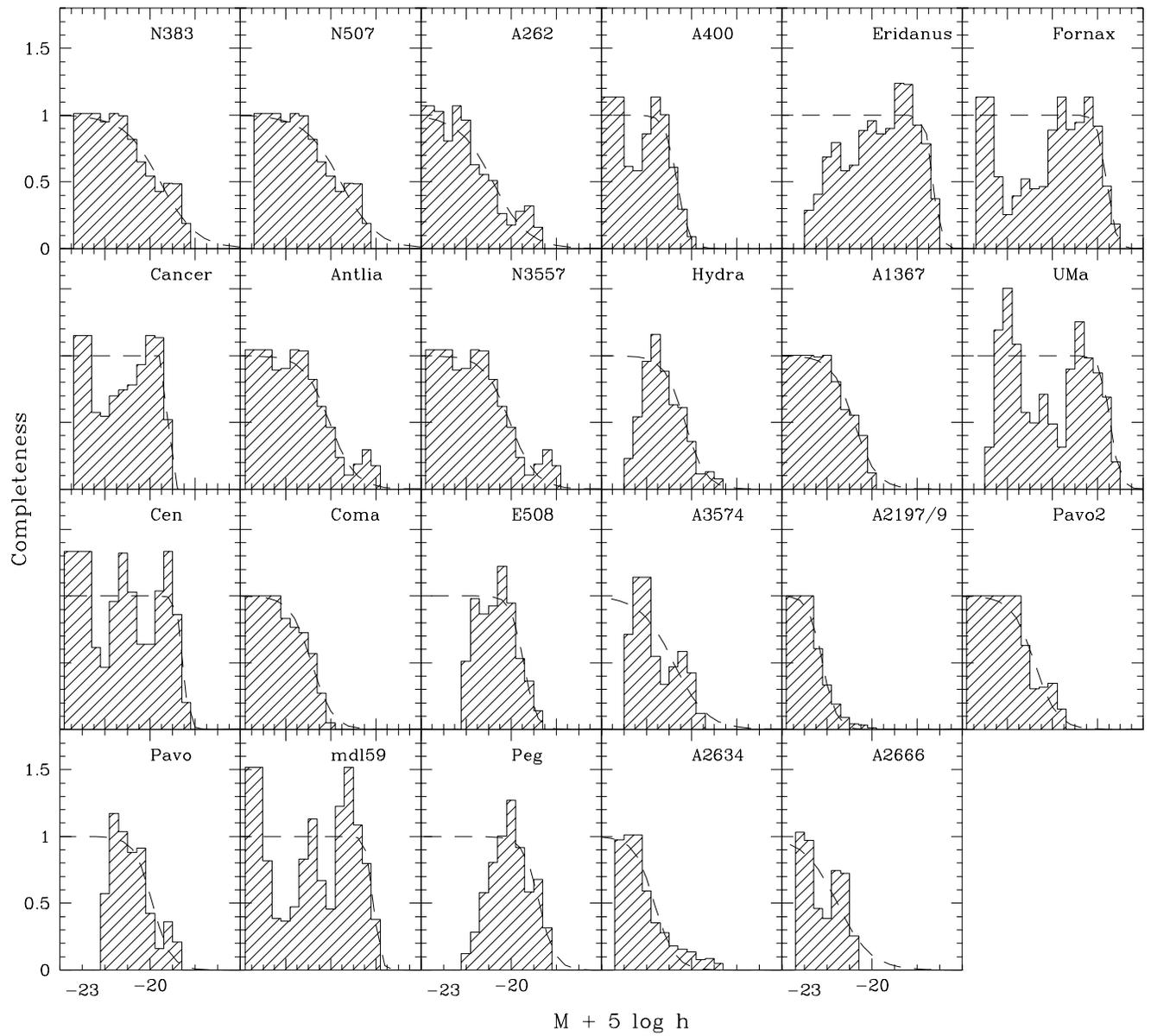
\begin{figure} 
\caption{Completeness histograms for each of the {\bf in+} cluster
samples. The shaded histograms are referred as the function $h'$ in section
3.2, while the dashed lines are the fitted completeness functions $c(y)$.
Parameters of the $c(y)$ functions are listed in Table 2.}
\end{figure}

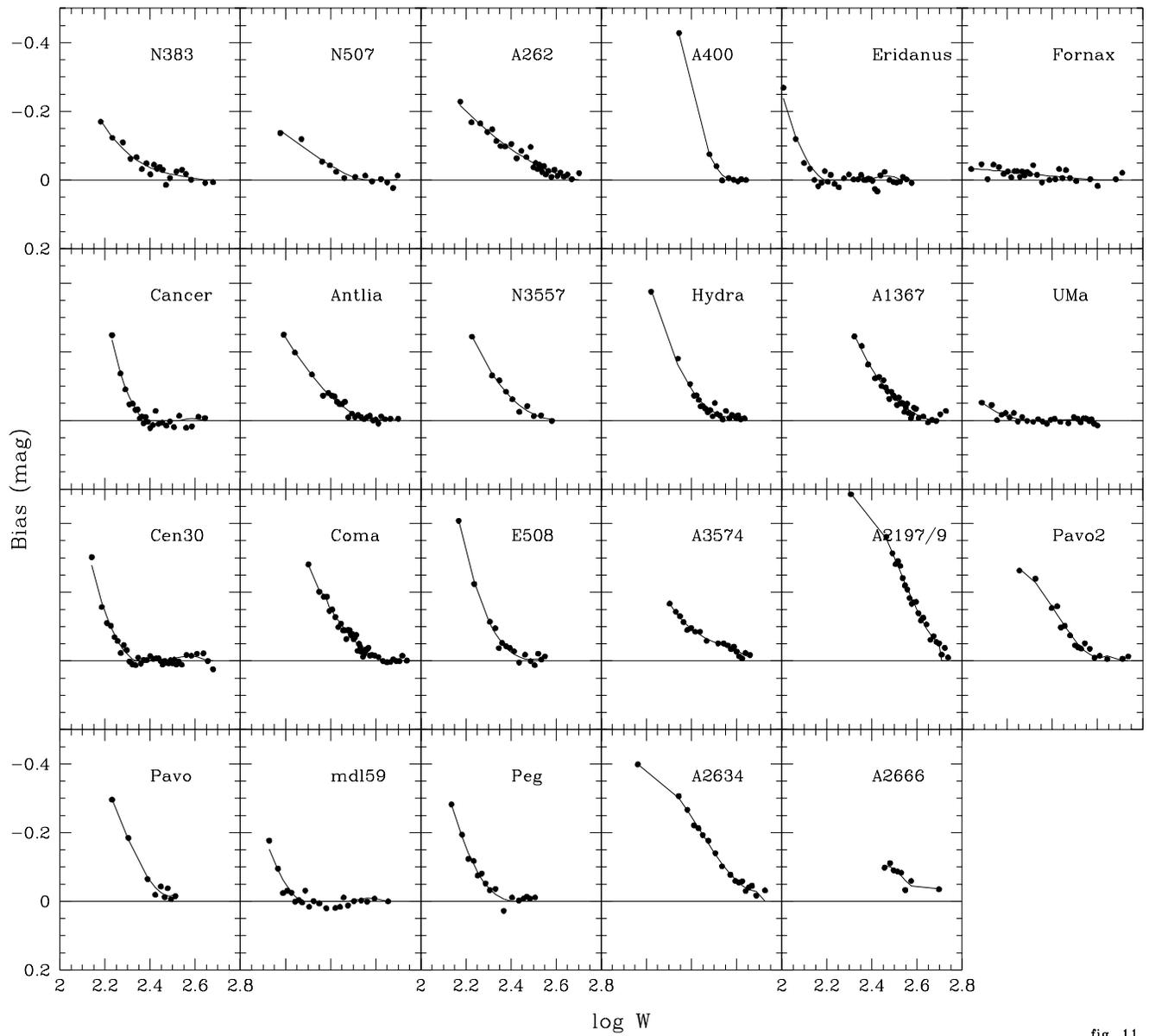
\begin{figure} 
\caption{Incompleteness bias, in magnitudes, computed for each {\bf in+}
cluster sample. Refer to section 3.2 for details.}
\end{figure}

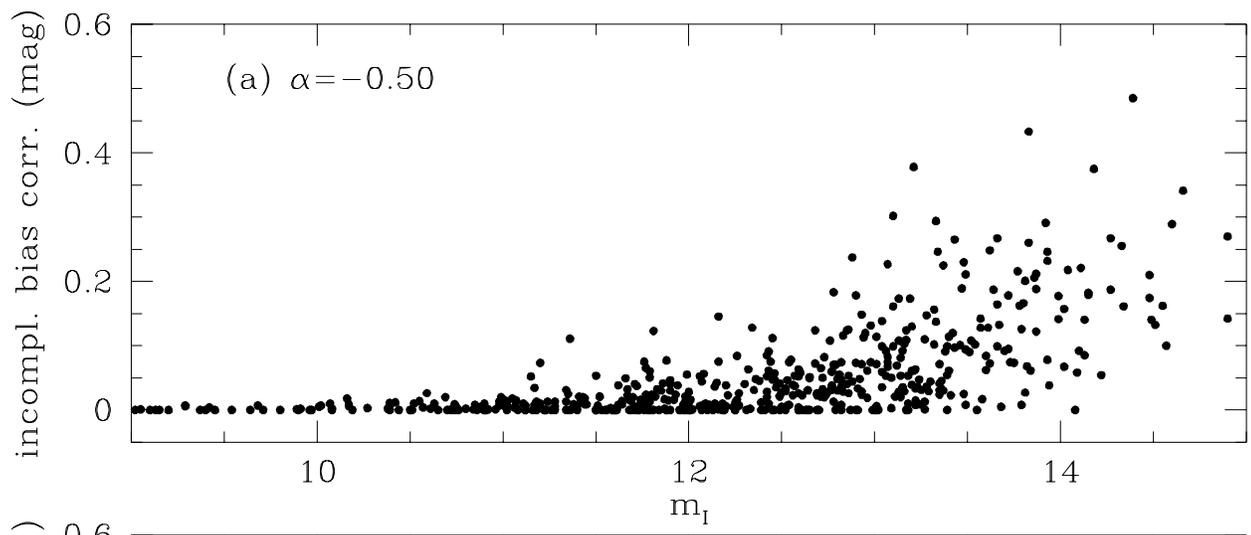
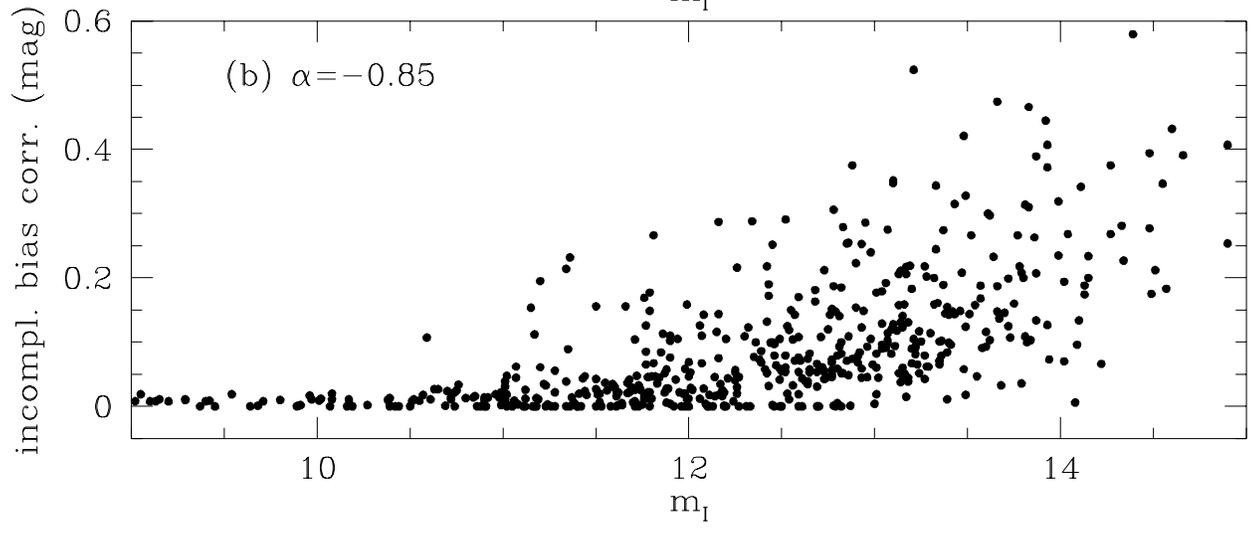
\begin{figure} 
\caption{Incompleteness bias correction applied to individual galaxies in the
{\bf in+} cluster samples, plotted versus apparent magnitude. In panel (a), the
bias correction is computed for a value of $\alpha=-0.50$ for the power law
exponent of the faint end of the LF, while in panel (b) a
value of $\alpha=-0.85$ is used.}
\end{figure}

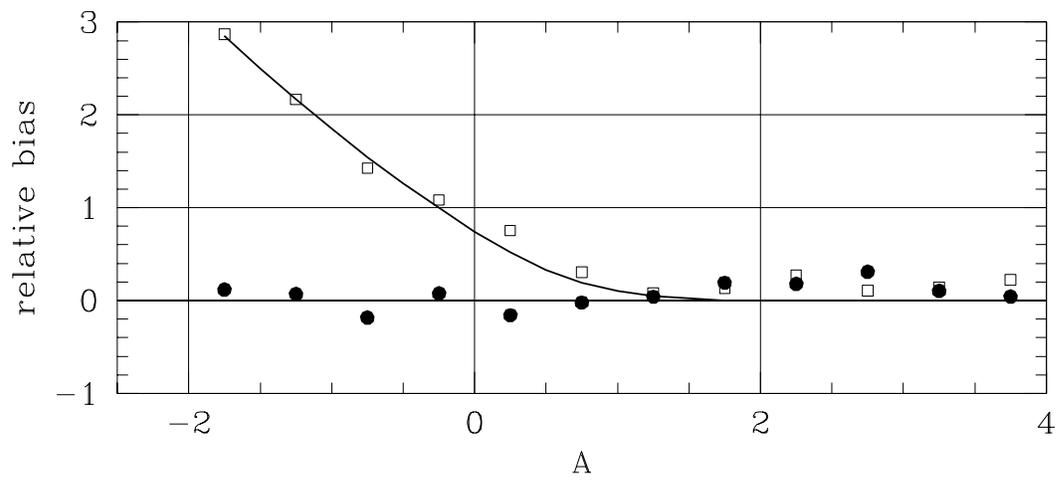
\begin{figure} 
\caption {The relative bias $B$, measured in units of the mean TF scatter, 
is plotted versus the parameter $A$, which is a measure of
the closeness to the edge of the magnitude limit of the catalog from which
sample objects are extracted. The ``catalog edge'' occurs at $A=0$. The values
of this bias were estimated for a strictly apparent magnitude limited sample
using Willick's (1994) approach (solid line), and our Monte Carlo approach
(unfilled squares). Both results agree well with each other. The solid symbols
display the bias expected for a sample with completeness characteristics
similar to those of our {\bf in+} global sample, estimated using a Monte
Carlo approach. See section 3.3 for details.}
\end{figure}

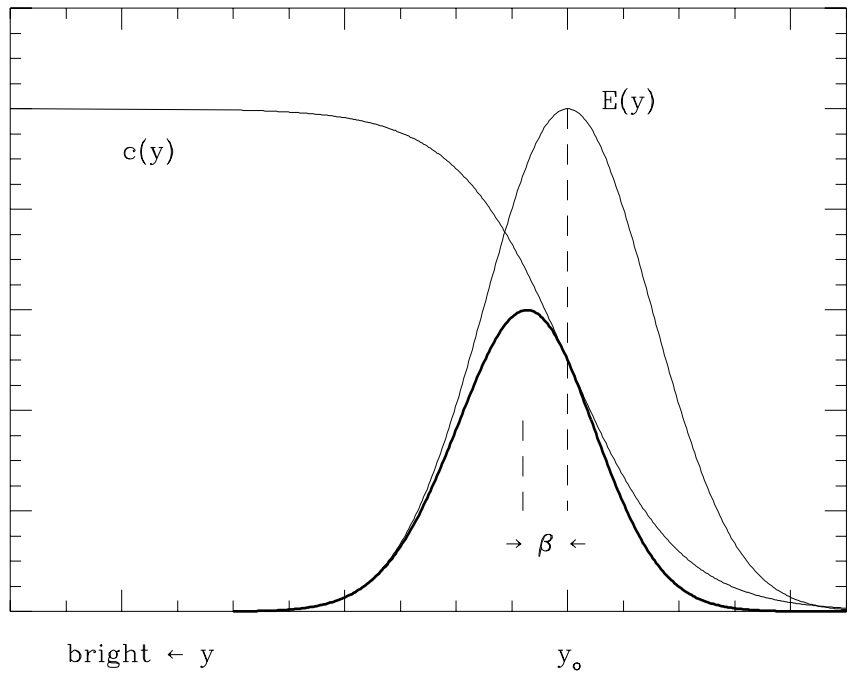
\begin{figure} 
\caption{Illustration of the cluster size--incompleteness bias.
The curve labelled $c(y)$ describes the completeness function of the
sample, which is complete on the left side of the plot. The curve
labelled $E(y)$ describes the distribution of magnitudes $y$ to be
expected from a population of galaxies of fixed magnitude $y_\circ$,
spread over the cluster; the spread reflects the depth of the cluster
along the line of sight. The heavy tracing is the product $c(y)E(y)$,
and the offset between the centroid of $cE$ and $y_\circ$ is the bias.}
\end{figure}

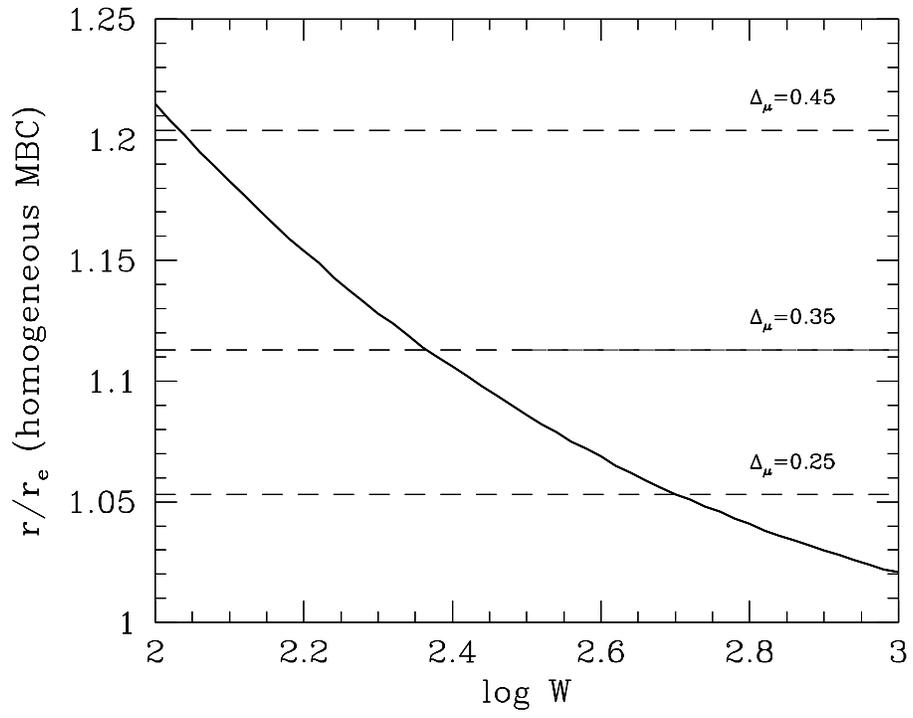
\begin{figure} 
\caption{Estimates of the homogeneous MBC. The solid line is the MBC
for a single object for which the distance modulus uncertainty is 
given by eqn. (3). By comparison, dashed lines display the MBC for
constant uncertainties on the distance modulus, as posted in mag.}
\end{figure}

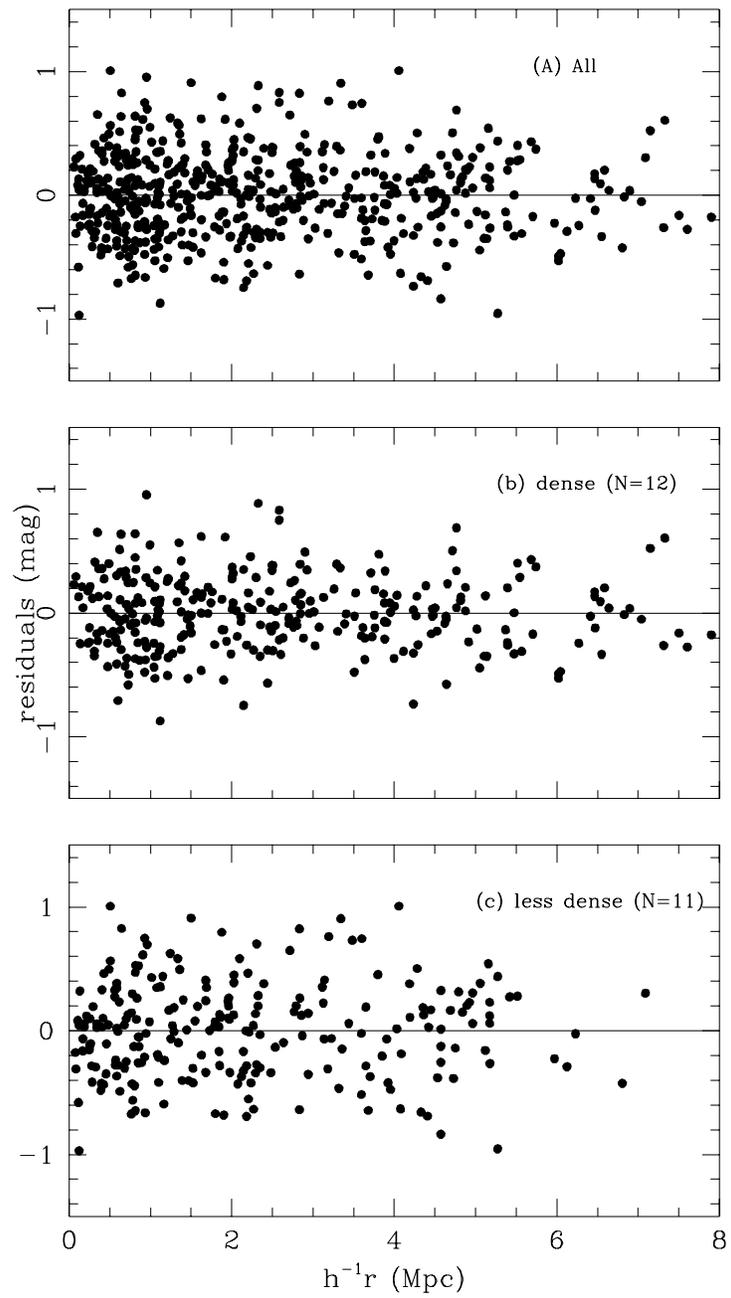
\begin{figure} 
\caption{TF magnitude residuals plotted versus projected radial
distance from cluster centers. In panel (a), all galaxies in the
{\bf in+} samples are included. In panel (b), only galaxies in 
the 12 higher density clusters (A262, A400, Cancer, Hydra, A1367,
Cen30, Coma, ESO508, A3574, A2197/9, A2634 and A2666) are plotted; 
in panel (c) galaxies in the remaining 11 clusters are in display.}
\end{figure}

\begin{figure} 
\caption{TF relations for individual cluster. Filled symbols
refer to galaxies in {\bf in} samples; additional objects in {\bf in+}
samples are plotted as unfilled symbols. The data are uncorrected for
any of the biases discussed in section 3. The unfilled symbols for
A2197 include peripheral cluster members as well as galaxies in A2199. The same fiducial line of slope -7.60 has been drawn in every graph. 
Absolute magnitudes are computed assuming the clusters are at rest 
with respect to the CMB reference frame, i.e. no correction for 
cluster motion is applied.}
\end{figure}

\begin{figure} 
\caption{TF relations for individual clusters. All galaxies in the
{\bf in+} samples are plotted; uncertainties $\epsilon_x$ and
$\epsilon_y$, arising from both measurement errors and from corrections
such as internal extinction, turbulence and disk inclination, are
plotted at each data point. No allowance for intrinsic scatter is made
in the error bars. The location of the data points is that obtained
after bias corrections discussed in section 3 have been applied. The 
same fiducial line of slope $-7.60$ has been drawn in every graph. 
Absolute magnitudes are computed assuming the clusters are at rest 
with respect to the CMB reference frame, i.e. no correction for 
cluster motion is applied. Data for the two clusters A2197 and A2199 
have been plotted in the same graph. For comparison, along the y--axis
of the panels on the left we show the assumed LF of the  
population from which each sample is estimated to be an incomplete representation.}
\end{figure}

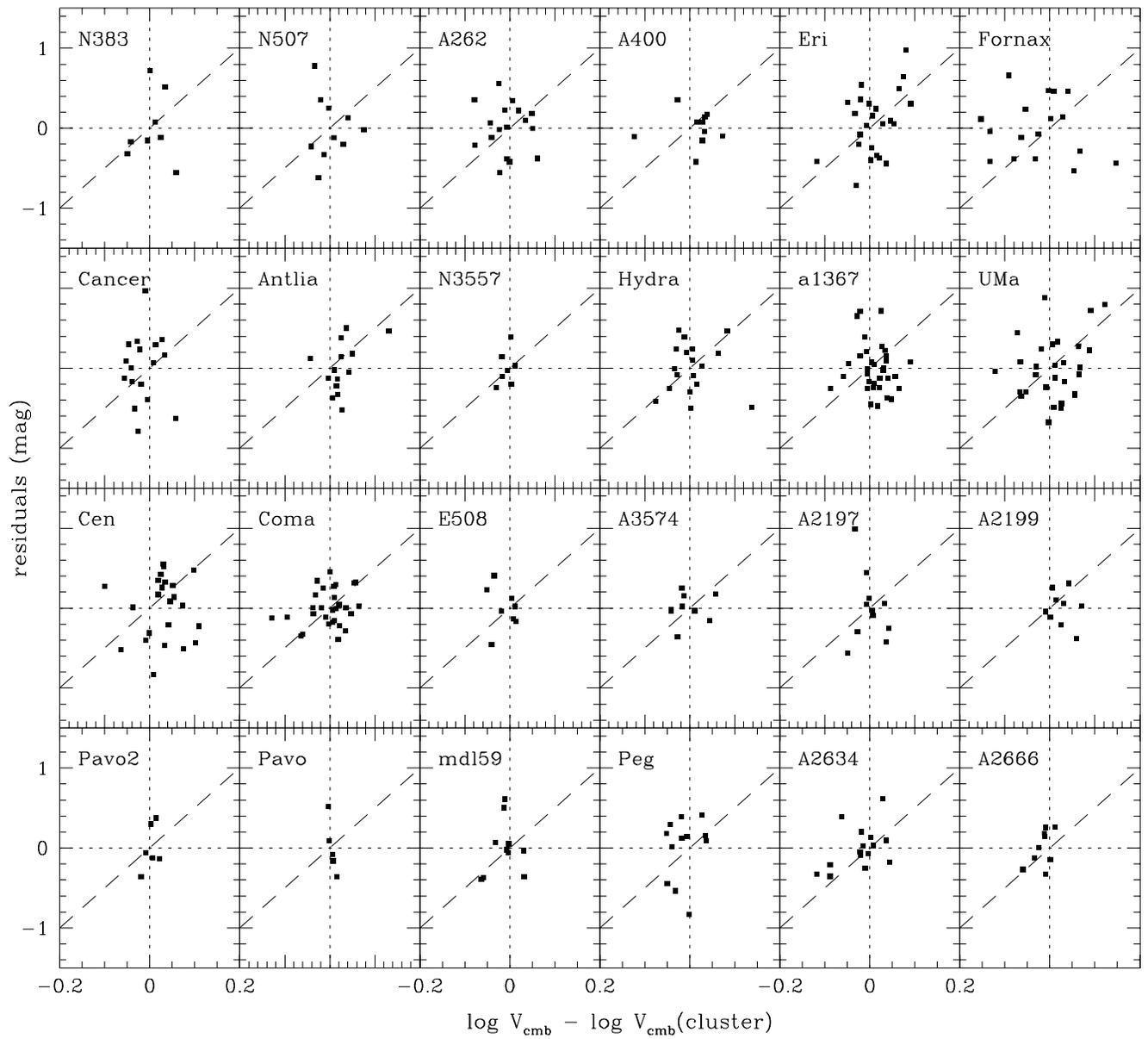
\begin{figure} 
\caption{Distribution of magnitude residuals for each {\bf in} cluster 
sample, about a TF relation of slope -7.62 and where the mean residual 
for each cluster is forced to be zero, plotted versus the log of each 
galaxy's radial velocity, minus the log of the cluster systemic radial 
velocity. A dashed line of slope 5 is drawn in each graph, indicating the 
locus about which galaxies that are not cluster members should scatter.}
\end{figure}

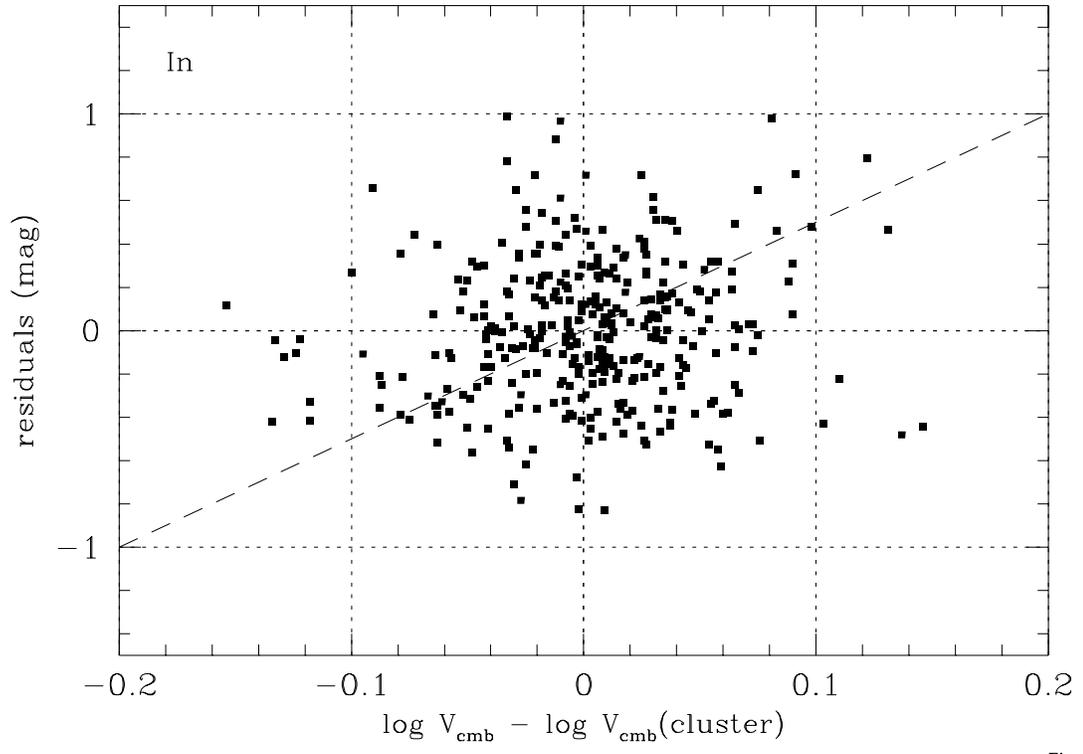
\begin{figure} 
\caption{Distribution of magnitude residuals for all galaxies 
in {\bf in} cluster samples displayed in Figure 19. A dashed line of 
slope 5 is drawn in each graph, indicating the locus about which 
galaxies that are not cluster members should scatter. There is no 
significant indication of sample contamination by foreground or 
background objects.}
\end{figure}

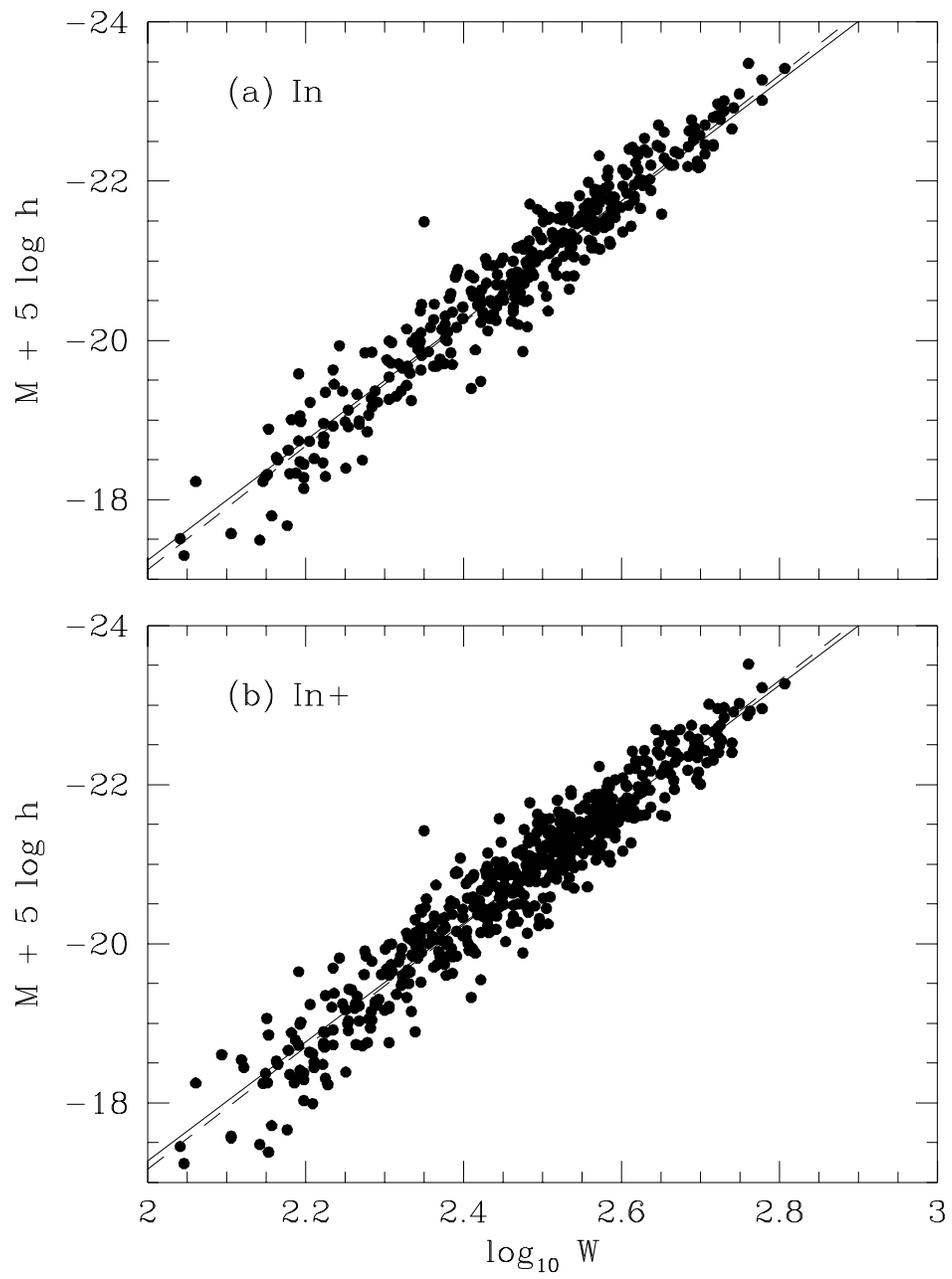
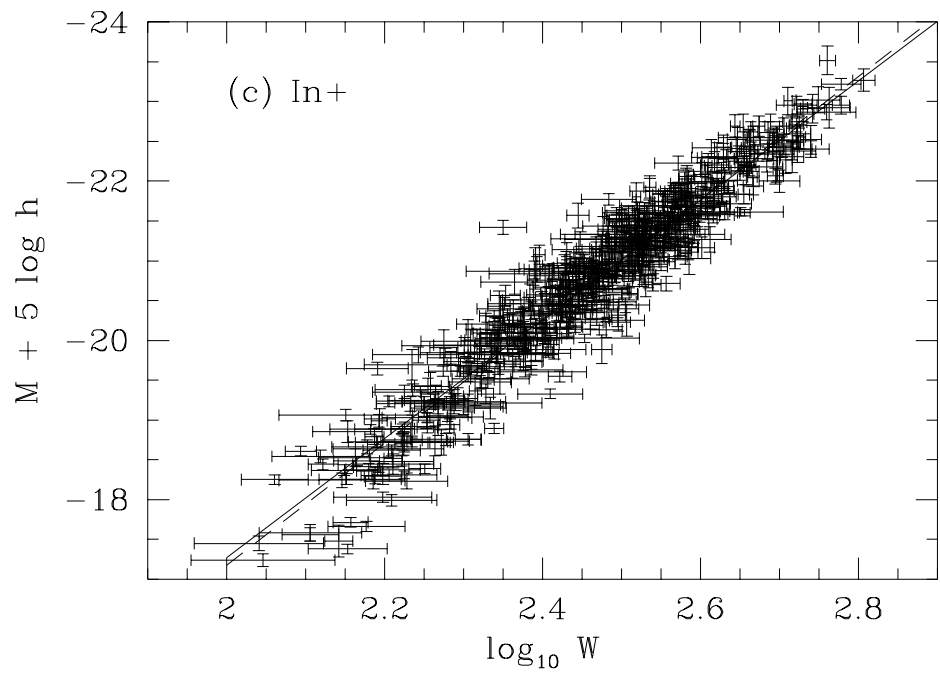
\begin{figure} 
\caption{Global set, obtained by combining the cluster samples
as described in section 6.1. Direct and bivariate regression fits
are inset. In panel (a), the result of the combination of {\bf in} samples
is shown; in panel (b), the analogous plot for {\bf in+} samples is displayed,
while in panel (c) we show the same data points as in panel (b), with their
associated measurement error bars.}
\end{figure}

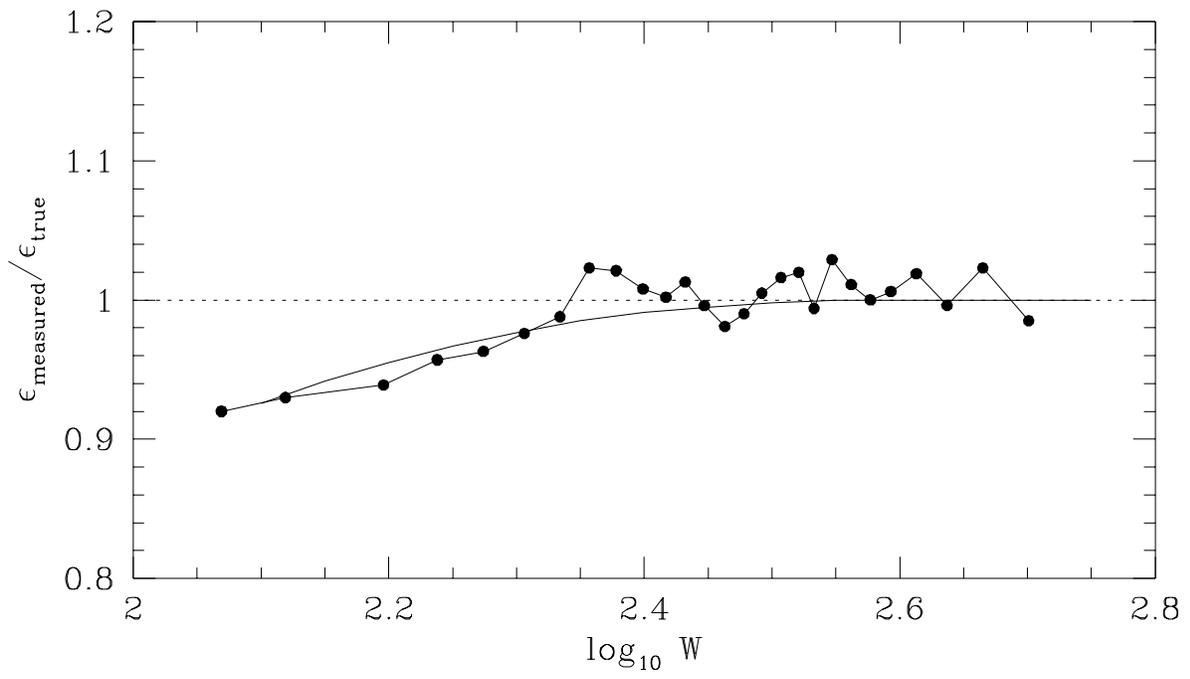
\begin{figure} 
\caption{Illustration of the underestimation of the TF scatter due to the
incompleteness bias for each {\bf in+} cluster sample, plotted
as a function of velocity width, for the the case of the global set 
combination shown in Figure 21(b). The smooth line superimposed to the data
points represents a parametrization of the scatter underestimation, which
was applied in the derivation of the position of the filled symbols in
Figure 1.}
\end{figure}

\newpage

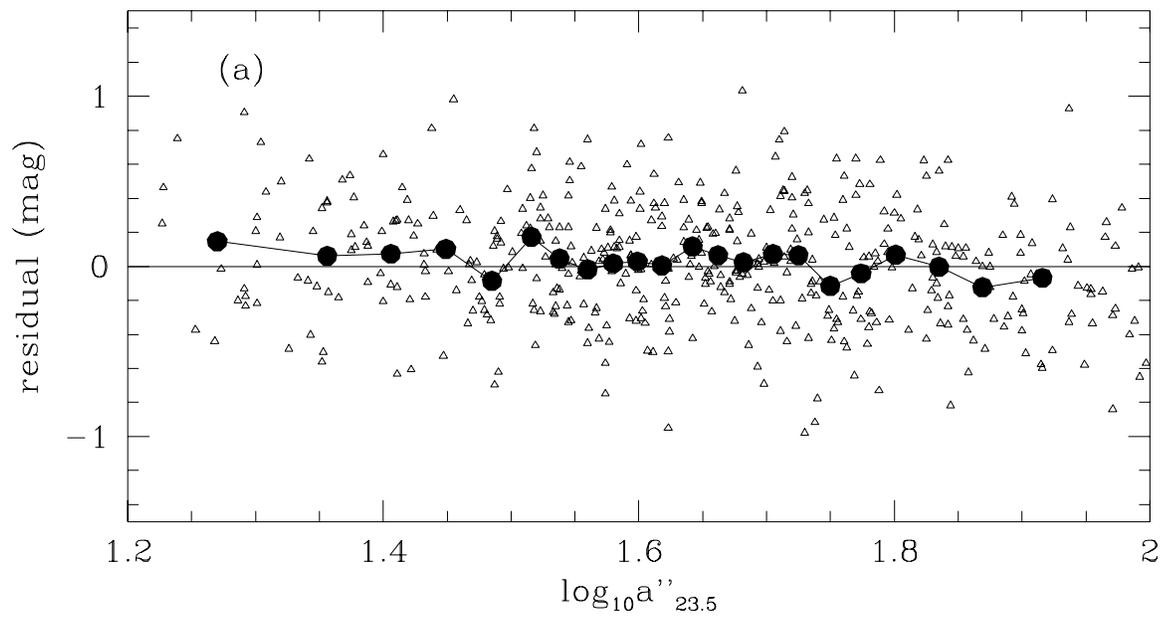
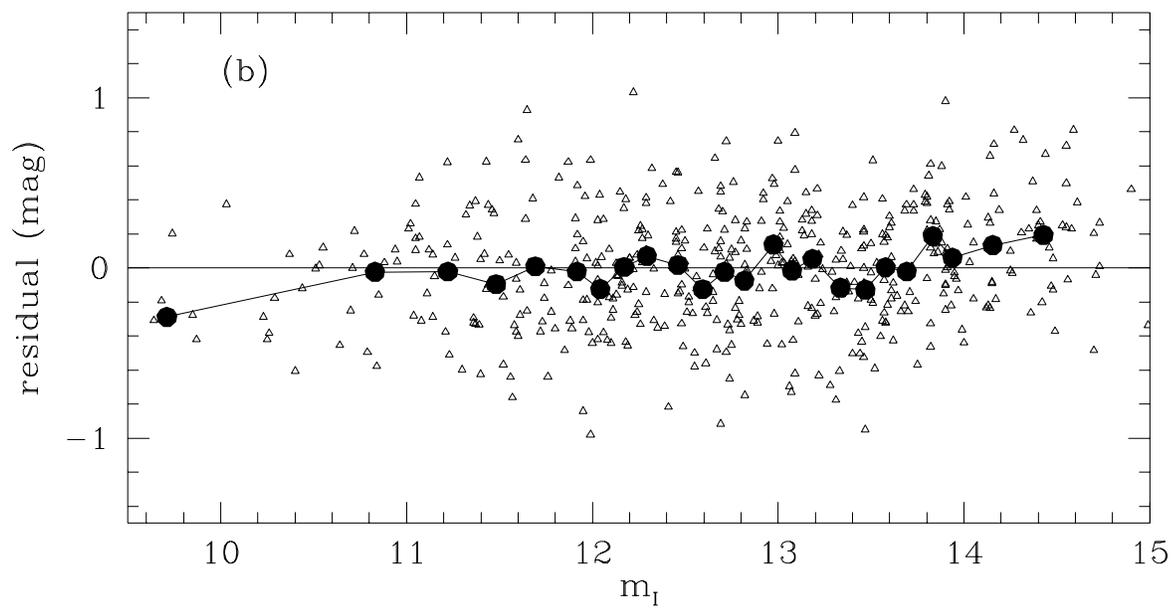
\begin{figure} 
\caption{(a) Magnitude residuals with respect to the bivariate TF relation
for the in+ global set, plotted versus isophotal angular diameter,
measured at the 23.5 mag arcsec$^{-2}$ isophote; (b) same residuals, plotted
versus apparent magnitude. In both cases, large solid symbols refer to
running averages.}
\end{figure}

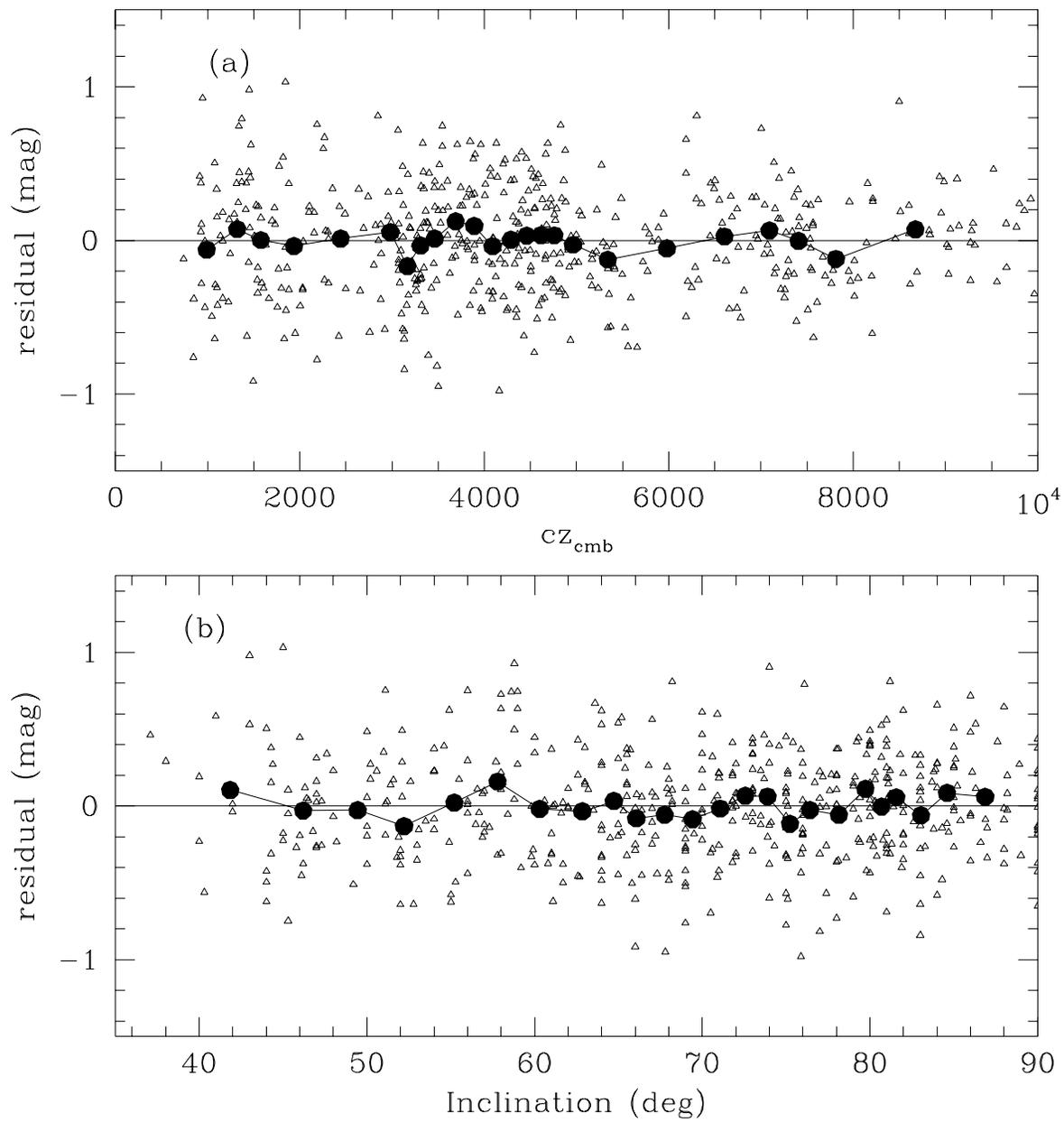
\begin{figure} 
\caption {(a) Magnitude residuals with respect to the bivariate TF relation
for the {\bf in+} global set, plotted versus galaxy redshift; (b) same residuals, 
plotted versus disk inclination. In both cases, large solid symbols refer to
runnign averages.}
\end{figure}

\end{document}